\documentclass[12pt, letterpaper]{article}

\usepackage{amssymb,amsmath,amsfonts,eurosym,geometry,ulem,graphicx,caption,color,setspace,sectsty,comment,footmisc,caption,pdflscape,subfigure,array,hyperref, bbm, tabularray, palatino}
\usepackage[sc]{mathpazo}
\usepackage[authoryear]{natbib}
\usepackage{longtable}
\usepackage{lscape}
\usepackage{indentfirst}
\usepackage[table,xcdraw]{xcolor}
\usepackage{fancyhdr}
\renewcommand{\footnoterule}{\vfill\kern -3pt \hrule width 0.4\columnwidth \kern 2.6pt}
\usepackage{titlesec}
\titleformat{\section}
{\normalfont\fontsize{16pt}{16}\bfseries}{\thesection}{1em}{}
\titleformat{\subsection}
{\normalfont\fontsize{14pt}{14}\bfseries}{\thesubsection}{1em}{}
\titleformat{\subsubsection}
{\normalfont\normalsize\bfseries}{\thesubsubsection}{1em}{}

\hypersetup{
    colorlinks,
    citecolor=blue,
    linkcolor=blue,
    backref=true
}

\newcolumntype{L}[1]{>{\raggedright\let\newline\\arraybackslash\hspace{0pt}}m{#1}}
\newcolumntype{C}[1]{>{\centering\let\newline\\arraybackslash\hspace{0pt}}m{#1}}
\newcolumntype{R}[1]{>{\raggedleft\let\newline\\arraybackslash\hspace{0pt}}m{#1}}
\newcommand\blfootnote[1]{%
  \begingroup
  \renewcommand\thefootnote{}\footnote{#1}%
  \addtocounter{footnote}{-1}%
  \endgroup
}

\begin{document}

\begin{titlepage}
\title{\textbf{From Transcripts to Insights: \\ 
Uncovering Corporate Risks Using\hspace{-0.1cm} Generative AI}}
\author{Alex G. Kim\thanks{The University of Chicago, Booth School of Business, alex.kim@chicagobooth.edu} \and Maximilian Muhn\thanks{The University of Chicago, Booth School of Business, maximilian.muhn@chicagobooth.edu} \and Valeri V. Nikolaev\thanks{The University of Chicago, Booth School of Business, valeri.nikolaev@chicagobooth.edu}}
\date{First Draft: October 5, 2023}
\maketitle
\blfootnote{\noindent We thank Saketh Aleti, Laurence van Lent, Yin Luo, Yinan Su, and workshop participants at the 7th Annual Global Quantitative and Macro Investment Conference by Wolfe Research for helpful comments. Irene Tan and Yijing Zhang provided excellent research assistance. This research is funded by the Fama-Miller Center for Research in Finance and the Stevens Doctoral Program at the University of Chicago Booth School of Business. Financial Support from the University of Chicago Booth School of Business is gratefully acknowledged.}

\begin{abstract} \noindent We explore the value of generative AI tools, such as ChatGPT, in helping investors uncover dimensions of corporate risk. We develop and validate firm-level measures of risk exposure to political, climate, and AI-related risks. Using the GPT 3.5 model to generate risk \textit{summaries} and \textit{assessments} from the context provided by earnings call transcripts, we show that GPT-based measures possess significant information content and outperform the existing risk measures in predicting (abnormal) firm-level volatility and firms' choices such as investment and innovation. Importantly, information in risk assessments dominates that in risk summaries, establishing the value of general AI knowledge. We also find that generative AI is effective at detecting emerging risks, such as AI risk, which has soared in recent quarters. Our measures perform well both within and outside the GPT's training window and are priced in equity markets. Taken together, an AI-based approach to risk measurement provides useful insights to users of corporate disclosures at a low cost.
\\  



\center{\textcolor{red}{\textbf{CAVEAT NOTICE:}}\\
The findings presented here are preliminary and subject to revision. We are currently reevaluating our data, methods, and conclusions.}

\bigskip
\end{abstract}
\setcounter{page}{0}
\thispagestyle{empty}
\end{titlepage}
\pagebreak \newpage

\pagestyle{fancy}
\fancyhf{}
\rhead{\thepage}
\chead{Uncovering Corporate Risks Using Generative AI}
\renewcommand{\headrulewidth}{0.5pt}

\begin{sloppypar}

\onehalfspacing

\section{Introduction} 
In the era of global political instability, climate uncertainty, and rapid technological change, corporations face multifaceted risks that extend far beyond traditional financial metrics. Among these are the emergent and swiftly evolving spheres of regulatory, environmental, and AI-related risks, each of which carries substantial implications for long-term growth and stakeholder value. This study aims to bridge the gap between generative AI technology and risk assessment methodologies by examining the potential of large language models (LLMs) to detect and analyze these critical aspects of corporate risk. By leveraging recent advances in language modeling, we seek to understand the capabilities of AI in navigating the complex corporate risk landscape and, ultimately, helping stakeholders to make more informed decisions in the face of growing uncertainties.

The evaluation of firm risks through textual analysis of corporate disclosures received substantial attention in recent literature \citep[see][for example]{hassan2019firm, hassan2021sources, chava2022measuring, sautner2023firm}. A distinctive feature of these studies is the utilization of dictionary-based bigram (n-gram) frequencies to quantify various risk types.\footnote{This approach counts the presence of risk-related words mentioned in the vicinity of risk-topic-specific bigrams from a pre-constructed dictionary (e.g., the algorithm searches for instances where the bigram ``economic policy" is used along with the word ``risk").} This literature has laid important groundwork that improves our understanding of corporate risks. However, the recent developments in AI technology provide a useful opportunity to delve deeper into textual data and extract a richer and more nuanced understanding of corporate risks. The new generation of language models is capable of understanding complex relationships within a text, incorporating the context within which relevant topics are discussed, and even making inferences. These aspects are critical for a comprehensive analysis of complex corporate risks. Furthermore, rapidly evolving changes in the political, environmental, and technological landscape in recent years quickly render existing dictionaries outdated or incomplete.

Two additional features make generative language models particularly attractive in the analysis of corporate risks. First, their general nature allows them to go beyond the context of a given text. Unlike traditional methods that analyze risks based on a single document, such as a conference call transcript, generative language models are trained on vast corpora that enable them to leverage \textit{general knowledge} acquired from similar documents or documents featuring related topics. This ``general AI" feature holds promise in improving the measurement of corporate risks because companies need not explicitly discuss risks in their disclosures.\footnote{For example, political risk is implicitly implied if an executive states that a firm is subject to a new regulation.} Second, an important advantage of the new generation of language models is that they synthesize the extracted information into coherent, understandable narratives, thus providing not only a quantitative assessment but also an explanation to support it.

To illustrate the usefulness of the AI-based approach, consider the example of SK Telecom Inc., a Korean cell phone service provider cross-listed and operating in the US. Figure \hyperref[fig1]{1} illustrates a significant wedge between the bigram-based measure of political risk, which implies no risk in 2018, and our GPT-based measure, which places the company towards the top of risk distribution. What is causing this difference? Around 2018, an active discussion took place worldwide regarding regulating the bundling of phone handset contracts and telecom services.\footnote{Indeed, Japan banned the bundling in 2019. See these two articles: \href{https://www.mobileworldlive.com/asia/asia-news/japan-steps-up-efforts-to-ban-device-subsidies/}{(link 1)} and \href{https://www.mobileworldlive.com/asia/asia-news/japan-bans-bundled-mobile-offerings/}{(link 2)}. Such bundling is considered detrimental for consumers since it makes pricing opaque \citep{hazlett2018mobile}.} During SK Telecom's 2018 earnings call, analysts actively asked questions concerning the bundled sales. The transcript, however, does not contain an explicit discussion of political or regulatory risks, and hence, these risks are not captured by bigrams. In contrast, GPT identifies "\textit{the separation of handsets and telecom services}" as a potential source of regulatory uncertainties and provides a rich explanation of this issue (see Appendix A). The example illustrates that LLMs can not only connect diverse pieces of information in a given text but also make judgments by interacting the context with the model's general knowledge. 

Despite this potential, LLMs' ability to evaluate firm-level risks is yet to be understood. In this study, we address this question by developing firm-level risk exposure measures using OpenAI's GPT3.5-Turbo LLM. In particular, we extract pertinent risk-related information from earnings conference call transcripts spanning January 2018 to March 2023. We then evaluate how GPT-based risk exposure measures compare to the existing measures in predicting stock market volatility and related economic outcomes.

We focus on the sources of corporate risks that are of the highest significance to firms' stakeholders: political risk \citep{hassan2019firm}, climate-related risk \citep{sautner2023firm}, and AI-related risk. Because AI risk is a new phenomenon, it enables us to probe the ability of language models to assess emerging risks. For each of the three measures, we induce additional variation by generating two types of output: (1) risk \textit{summaries} and (2) risk \textit{assessments}, as discussed next. 

For risk summaries, we specifically instruct GPT to focus solely on the document contents and avoid making judgments. These instructions minimize GPT's reliance on its general knowledge. The risk summaries are thus human-readable reorganizations of risk-related discussions in conference call transcripts. In contrast, risk assessments utilize the unique ability of LLMs to integrate the documents' context with their general knowledge and make judgments. In this task, we develop a prompt that instructs GPT to generate an assessment of a given risk, which is not limited to the information included in the given transcript. 

There are a number of possible approaches to converting readable summaries into quantitative risk exposures. One could, for example, rank the summaries based on their contents or train a model to construct risk scores predictive of future uncertainty. While such approaches are likely to add value, we follow a simple approach that computes the ratio of the length of risk summaries (assessments) to the length of the entire transcript. Higher ratios are interpreted as higher risk exposure.

Our measures show intuitive variation. The tobacco industry exhibits the highest political risk, coal mining exhibits the highest climate risk, and business services exhibit the highest AI risk exposures. The aggregate time series of our political and climate risks covary positively with the corresponding bigram-based exposure measures, culminating in the first quarter of 2020.\footnote{This spike is likely to be attributable to the presidential election of the US and the outbreak of COVID-19. As the Biden administration is known to put more emphasis on global environmental issues than the previous administration, we see an elevated level of climate change risk after the first quarter of 2020. Furthermore, in line with the Russia-Ukraine war, we also observe an increase in political risk (and also climate change risk) in the first quarter of 2022.}$^,$\footnote{For the AI-related risk exposure measures, there is no corresponding bigram-based measure.} AI-related risks exhibit a different pattern. Instead of an uptick around 2020, we observe a sharp increase in the most recent years, consistent with the emergence of AI technologies. 

Variance decomposition analysis shows that time- and industry-time-invariant characteristics can only explain 10-15\% of the total variation across the three risk measures. In line with \cite{hassan2019firm}, about 90\% of variance in political risk is firm-specific. This number goes down somewhat (is similar) for climate risk (AI risk) proxies. We further decompose firm-level residual variation and find that firm-invariant factors explain only 20-30\%. Thus, time-varying firm-specific components account for the bulk of the variation in firm-specific exposures to political, climate, and AI risks.

Our main analysis uses a market-based approach to evaluate whether GPT-based proxies are effective at measuring firm-level risks. Specifically, we examine whether risk exposure measures explain future stock price volatility \citep{engle2004risk}. We use two forward-looking firm-level volatility metrics: implied volatility derived from option prices \citep{hassan2019firm, sautner2023firm}, and abnormal realized volatility that builds on \citet{loughran2014measuring}. 

We begin by examining political and climate risk exposures as they received the most attention in prior literature. Focusing on the main sample period from 2018 to 2021, we show a strong and robust positive relation between the GPT-based risk exposure measures and stock price volatility. Across different fixed effects structures, GPT-based political and climate risk measures exhibit positive associations with both of our volatility proxies. We also consistently find evidence that GPT-based proxies are more informative compared to the bigram-based proxies in explaining stock price volatility for these two types of risks, indicating the significant value derived from the new technology. In particular, GPT-based measures subsume information in bigram-based measures. More importantly, when comparing our GPT-based measures against each other, risk assessments perform better than risk summaries both in the case of political and climate risks. This result implies that AI-generated insights are useful in uncovering corporate risks. 

We also examine our risk proxies using the time period unseen by the language model during its training phase (pure out-of-sample period). Since GPT3.5 is trained on texts that precede September 2021, we focus on a sample of transcripts from January 2022 until March 2023 for additional tests. Even within this limited sample, we find strong and robust evidence that GPT-based risk exposure measures are positively associated with volatility variables, suggesting that our results are unaffected by GPT's possible ex-post knowledge. 

Turning to the analysis of AI-related risks, based on our main sample, we find limited evidence that AI risk is predictive of stock market volatility. It shows up only in one of the two models featuring risk assessment-based proxies. This result is not unexpected, given the recency of AI disruptions \citep{eisfeldt2023generative}. Indeed, we show that the AI-related risk proxy becomes significant in explaining volatility in the most recent two years.

Having established the validity of the GPT-based risk proxies, we turn our attention to firms' actions predicted to change as a result of each risk exposure. First, we investigate whether our risk measures can explain capital investments.\footnote{We use recursive capital expenditure following \citet{brink2017measuring} and find that climate change risk exposure and political risk exposure are associated with lower investments.} In theory, riskier companies experience higher financing costs and value the option of waiting \citep{dixit1994investment}. Thus, ceteris paribus, they are less likely to make investments. This force is expected to be pronounced for political and environmental risks. However, for technology-related risks, the effect is less clear because addressing AI challenges requires significant investments in new technology. Indeed, the political and (less so) climate-risk exposures exhibit negative associations with investment, whereas AI risk exhibits a positive albeit insignificant relation. Furthermore, the positive effect of AI risk becomes significant during the 2022-2023 period.

We find that firms further adjust their behavior in response to specific risks they are facing. They increase (1) lobbying activity in response to political risk, (2) green patent filings in response to climate risk, and (3) AI-related patent filings in response to AI risk. These findings continue to hold in the period outside of GPT's training window.

Our additional analysis indicates that while political and climate risks vary up and down in their importance over our sample period, which includes the Covid pandemic, the importance of AI-related risk has been steadily growing. Finally, we show that environmental and AI risks command significant equity risk premia when assessed based on the traditional asset pricing methodology. 

We make the following contributions to the literature. First, we probe the economic usefulness of AI-powered large language models in risk assessment. Although generative LLMs have much potential for assisting investors in analyzing complex, unstructured information, their economic usefulness in risk assessment and risk management is yet to be understood. We contribute to a nascent and actively developing body of work on the value of LLMs \citep[see][]{bernardblankespoor2023, lopez2023can, jha2023chatgpt, eisfeldt2023generative, kim2023bloated, Chen2023expected}, by showing that AI tools are effective at distilling disclosures to extract information about diverse risk categories.\footnote{Generative AI tools are effective at generating informative summaries of corporate disclosures \citep{kim2023bloated}. They can also assess corporate policies \citep{jha2023chatgpt}, innovation success \citep{yang2023predictive}, and job substitutability \citep{eisfeldt2023generative}.}

Second, we contribute to the recent literature that uses corporate disclosures to construct firm-level measures of risk exposure: political risk \citep{hassan2019firm}, country risk \citep{hassan2021sources}, climate risk \citep{sautner2023firm}, inflation risk \citep{chava2022measuring}, and pandemic risk \citep{hassan2020firm}. We complement and build on this influential work by adopting AI-based technology to analyze risks. In contrast to existing studies that rely on topic-based bigram dictionaries, LLMs' are trained to understand the deeper context in which bigrams are encountered.\footnote{\citet{chava2022measuring} employ a deep-learning approach to classify the topic of each sentence and measure firm-level inflation exposure.} Indeed, we document that GPT-based measures are more informative than bigram-based measures and generally subsume their information content. 

Last but not least, we contribute by establishing the value of general AI for understanding complex topics like risk. We show that LLMs successfully leverage their general knowledge to derive \textit{insights} about corporate risks from a given context. These insights go beyond the information discussed in the processed document. The AI-knowledge-based evaluations of risks, unrestricted to the document context, are incrementally informative and generally outperform summaries that are based on localized knowledge. 

\section{LLMs' Theoretical Usefulness in Measuring Risk Exposure}
The recent emergence of Large Language Models (LLMs) has fundamentally transformed our ability to understand and generate text. AI tools that rely on these models, such as ChatGPT, have demonstrated exceptional abilities across various domains, from natural language processing to content creation. In this section, we discuss the foundations that make LLMs particularly well-suited for the analysis of multifaceted corporate risks. Leveraging the vast knowledge embedded within these models, they can be used to uncover valuable insights into the intricate corporate risk landscape. Indeed, corporate risk exposures are often subtly implied in conference call discussions rather than explicitly stated. Evaluating these exposures requires bridging the information found in call transcripts with users' prior knowledge, and LLMs have the potential to be a transformative force in this task. We delve into why LLMs outshine traditional linguistic approaches and offer unique advantages for extracting corporate risk insights from unstructured text data.

\subsection{How does an LLM encode textual information?}
Generative AI models, such as GPT by OpenAI and LLaMa by Meta, are deep neural networks trained on a large corpus of text data with the purpose of predicting the next word in a sentence within a larger text context. These models make use of word embeddings (high dimensional word vectors that encode word meaning) and rely on the revolutionary Transformer architecture \citep{vaswani2017attention}, trained to transform the meaning of words depending on their context (e.g., position in the sentence). Transformer employs a so-called attention mechanism, which effectively directs the model to focus attention on relevant words. For example, sentences that appear early in the call may be relevant when encoding the meaning of words later on. This ability to recognize related words is instrumental in analyzing sparsely distributed (risk-related) information in a long text sequence.\footnote{The transformer consists of multiple layers stacked on top of each other. As information progresses through these layers, the model refines its understanding of the content, allowing for more complex and nuanced interpretations.} 

When a user prompts a query requesting to summarize or assess risk exposure, LLMs choose which words and, consequently, sentences are most relevant to the query without confining the analysis to the vicinity of specific words (encompassing the entire text). In contrast, dictionary-based algorithms focus on specific bigrams and typically employ a relatively narrow "attention" window to determine their relevance. For example, bigrams that encode political or regulatory topics (e.g., environmental regulation) are counted as long as they are within a ten-word window relative to the word "risk." Furthermore, many companies' executives may avoid the explicit mention of "risk" or related words while, at the same time, a mention of, e.g., environmental regulation or similar topics implies an important form of uncertainty for the company. 

\subsection{The Value of General Knowledge}
Besides a context-based interpretation of the (risk-related) text, a key feature that distinguishes LLMs is their extensive general knowledge that facilitates the model's logical reasoning. Because the models are pre-trained on vast amounts of textual data and feature billions of parameters, in addition to learning language structure \citep{klafka2020spying}, they also acquire massive general knowledge \citep{bubeck2023sparks} and reasoning abilities \citep{liu2023evaluating, binz2023using}. This general knowledge forms a "prior" that the model uses when presented with a new text (context), and it affects how the model interprets the new data.

Fundamentally, the presence of extensive general knowledge allows the model not only to effectively summarize risk-related content but also to make its own assessment or judgment of risks based on the context provided by the conference call transcript. Such assessment is thus a blend of the model's knowledge and the context provided by the document. The Transformer's architecture enables this integration of knowledge at every step (layer), allowing it to generate coherent and contextually relevant insights.

\section{Methodology and Implementation}
Earnings calls are an attractive source of new information about corporate risks because they contain an informational exchange between the demand (analysts representing stakeholders) and the supply (executives) sides. This choice follows a number of prior studies measuring firm-level risk exposures based on earnings call transcripts \citep{hassan2019firm,hassan2021sources,sautner2023firm,chava2022measuring,hassan2020firm}. In contrast to these studies, we employ AI techniques to uncover risk exposures. Specifically, we use OpenAI's GPT3.5-Turbo LLM to analyze the transcripts. 

A visual representation of our GPT processing pipeline is provided in Figure \hyperref[fig2]{2}. GPT limits combined (input and output) tokens to 4,000, whereas call transcripts are 7,000 tokens on average. Thus, we chunk transcripts into several parts, which is a common practice in the literature \citep[see][for example]{goyal2022news, zhang2021exploratory, kim2023bloated}. Chunking can improve the quality of GPT's output because the model struggles to generate detailed summaries when processing long documents, whereas it produces summaries on par with humans when processing shorter essays \citep{choi2023chatgpt}. Chunking also significantly reduces the computational burden of calculating self-attention scores, which increases quadratically in document length.\footnote{For example, consider two documents A and B, with similar complexity. Assuming all other things equal, A has 10 tokens and B has 100 tokens. When calculating positional encodings and self-attention, the model has to consider 100 (=$10^2$) pairwise relations. However, when it comes to document B, the model needs to consider 10,000 (=$100^2$) pairwise relations.} 

To ensure the model's flexibility, we allocate 2,000 tokens for the input text and the remainder for the output text. To maximize chunk quality, we divide the transcript into a presentation and Q\&A sections. In the presentation part, we avoid splitting the speech by the same executive into different chunks. Similarly, we do not split Q\&A in the middle of the answer to an analyst's question.\footnote{When a question from an analyst prompts responses from multiple executives, we consolidate these answers into a single chunk to maintain coherence. Very occasionally, chunks may exceed the assigned input token limit. In such cases, we further divide the chunk into smaller chunks. Further chunking accounts for less than 3\% of the total GPT processing.} We concatenate the model's output at the call level (pooling across chunks).\footnote{We do not apply a summarization layer when concatenating the output texts, although this is an option since our purpose is to measure the degree of risk exposure. Even though some contents are repeated in different chunks, we interpret this as the topic (risk) being important for the firm. Similarly, bigram-based measures count the total number of occurrences rather than isolating the count of unique mentions.} In the absence of risk-related information, we instruct GPT to print "NA" and subsequently purge NAs from the output.\footnote{Doing so ensures zero values for the exposure measures when calls have no risk-related information.} 

We design separate prompts for risk assessments and risk summaries. For summaries, we instruct the model to ignore external information sources. For assessments, in contrast, the model is instructed to make judgments accompanied by narrative reasoning. In both cases, we provide the model with a context specifying that the input text is excerpted from an earnings call transcript. The prompt also provides an explanation of each risk and a list of sample questions that are relevant for understanding risk exposures.

The risk exposures corresponding to summaries, $RiskSum$, and assessments, $RiskAssess$, are constructed as follows: 
\begin{align}
    RiskSum_{it} & = \frac{\sum_{l=1}^{K_{it}} \text{len}(\textbf{S}(c_{it}^l))}{\text{len}(c_{it})} \\
    RiskAssess_{it} & = \frac{\sum_{l=1}^{K_{it}} \text{len}(\textbf{A}(c_{it}^l))}{\text{len}(c_{it})} 
\end{align}
where $c_{it}$ is earnings call transcript for a company $i$ in quarter $t$ divided into $K_{it}$ chunks $c_{it}^1, c_{it}^2, \cdots, c_{it}^{K_{it}}$. $\textbf{S}(\cdot)$ is a GPT-based function that generates risk summaries and $\textbf{A}(\cdot)$ is the corresponding risk assessment function. $\text{len}(\cdot)$ measures the number of words in a given text.\footnote{Note that $\text{len}(c_{it})$ might not equal $\sum_{l=1}^{K_{it}} \text{len}(c_{it}^l)$ since we drop several chunks such as operator instructions.} We set the temperature parameter for the text generator to zero and do not restrict the maximum output length.\footnote{High temperature might generate creative yet less replicable answers. While low-temperature values are generally appropriate, we set the temperature to zero (its minimum value) to keep the summaries as close as possible to the actual content of the transcript.} 

Our prompts are designed to capture three different types of \textit{RiskSum} and \textit{RiskAssess}. Specifically, we measure (i) political risk summary, \textit{PRiskSum}, and assessment, \textit{PRiskAssess}, (ii) climate risk summary, \textit{CRiskSum}, and assessment, \textit{CRiskAssess}, and (iii) AI risk summary, \textit{AIRiskSum}, and assessment, \textit{AIRiskAssess}. For political risk, GPT's output generally includes explanations of political or regulatory uncertainties, e.g., whether the company is likely to be affected by a new regulation. For climate risks, GPT's answers typically encompass how the company's operations might be impacted by extreme weather or environmental policy changes. Finally, for AI risks imposed by AI itself, GPT often discusses how a firm's primary operations might be replaced or assisted by AI, as well as whether the company's business is dependent on AI technologies.

\section{Data}

\subsection{Earnings Call Transcripts}
We source quarterly earnings call transcripts from Capital IQ S\&P Global Transcript database. Our analysis focuses on US firms' transcripts available between January 2018 and March 2023. We selected this period because (1) generating risk summaries and assessments for each risk metric is costly and time-consuming, (2) a considerable part of the sample is outside of GPT's training window (allowing for pure out-of-sample tests), and (3) this time period is characterized by significant changes in political, climate and AI uncertainty. 

A typical earnings call, which lasts approximately an hour, contains two parts. A presentation session, during which executives describe the company's performance, and a Q\&A session, during which analysts ask questions. Our data identifies portions of the transcript attributable to a given speaker, allowing us to distinguish between presentation and discussion sessions and between questions by different analysts.

As our analysis focuses on capital market outcomes, we restrict our sample to publicly traded companies. We exclude very short calls and calls without a discussion session \citep{brochet2016capital, baik2023managers}.\footnote{Such calls account for 1.4\% of the total transcript sample. We also drop operator instructions and chunks shorter than 50 tokens as they are more likely to be greetings or irrelevant conversations.} We exclude calls that are conducted in languages other than English, including the ones that are machine-translated into English. After applying these filters, our final sample consists of 69,969 transcripts from 4,983 distinct firms. We further split the sample into two periods: our baseline tests use calls from January 2018 to December 2021, and our post-GPT-training sample includes calls from January 2022 to March 2023.\footnote{Our benchmarks are bigram-based risk exposure measures from \citet{hassan2019firm} and \citet{sautner2023firm}. Their sample of released risk exposure measures ends in early 2022, and we match our sample period to theirs.}

\subsection{Capital Market Variables}
Our primary validity test for firm-level risk measures is based on the association between these risk measures and stock price volatility \citep{engle2004risk}. Specifically, we use two forward-looking volatility proxies: implied volatility and abnormal volatility, discussed next. Stock return data is sourced from the Center for Research in Security Prices (CRSP), and market returns and risk-free rates are from Ken French's website. 

\textit{\textbf{Implied Volatility}}. We obtain firm-quarter implied volatility from OptionMetrics, following \citet{hassan2019firm} and \citet{sautner2023firm}. OptionMetrics calculates implied volatility based on the Black-Scholes model for European options and the Cox-Ross-Rubinstein model for American options. We use the implied volatility derived from the 90-day at-the-money options measured as of the end of each fiscal quarter.

\textbf{\textit{Abnormal Volatility}}. Since implied volatility hinges on option models' assumptions, which may not reflect the most recent risk-related information (it is also limited to stocks with actively traded options), we also use realized \textit{abnormal} volatility. Following \citet{loughran2014measuring}, we measure volatility as the root mean squared errors (RMSE) from the market model residuals. Abnormal volatility is the ratio of post-conference call RMSE to pre-call RMSE calculated as follows. Post-call RMSE is from a market model estimated over the period starting six and ending 28 trading days after the conference call. Specifically, the market model is: $r_{it} = \beta_0 + \beta_1 r_{mt} + \varepsilon_{it}$, where $r_{it}$ is the daily stock return for company $i$ on day $t$ and $r_{mt}$ is the market return on day $t$. Similarly, using returns data from $-257$ to $-6$ days relative to the conference call date, we estimate pre-conference call RMSE.\footnote{We require at least ten valid observations to estimate the post-conference RMSE, and at least 60 valid observations for pre-conference RMSE. Additionally, in untabulated analyses, we also use alternative estimation windows for post-conference call RMSE estimation and obtained quantitatively and qualitatively similar results.} We then take the ratio of the two RMSE values to calculate the abnormal volatility: $$\text{Abnormal Volatility} = \frac{\text{Post-Conference RMSE}}{\text{Pre-Conference RMSE}}-1.$$

\subsection{Economic Variables}
Besides focusing on market-based outcomes, we also study firms' responses to risks by examining their real activities, discussed next. 

\textit{\textbf{Investments}}. We follow \citet{stein2013effect} to construct a capital investment measure using the ratio of current quarter capital expenditures ($CapEx_t$) to the recursively updated cumulative capital taken as of the previous quarter end ($K_{t-1}$). For the initial time period in our sample, $K_{1}$ is set to the quarter-end value of property, plant, and equipment from the Compustat Quarterly database. Subsequently, capital is recursively updated as follows: $K_{t} = K_{t-1}\times (1-\delta) \times (1+\rho_t)+ CapEx_t$, where $\delta$ is the depreciation rate assumed to be 10\%, and $\rho_t$ is the inflation rate measured by the change in the monthly Producer Price Index published by the Federal Reserve Economic Data (FRED). Accordingly, capital investment in period $t$ is measured as $\frac{CapEx_t}{K_{t-1}}$.

\textbf{\textit{Lobbying Activity}}. The Lobbying Disclosure Act of 1995 mandates that each lobbying firm reports its lobbying expenditures and their recipients. We obtain lobbying data from the Center for Responsive Politics, which maintains an archive of lobbying records (as long as lobbying amounts exceed the mandated disclosure threshold). We fuzzy match the names disclosed in lobbying reports to firm names on Compustat. Because lobbying amounts are highly skewed, we use a lobbying indicator $\textbf{1}\left(\text{\$ Lobby Amount}>0\right)$, which takes the value of one when a firm engages in any lobbying, and zero otherwise.

\textbf{\textit{Green Patents and AI-Related Patents}}. We obtain patent data from the United States Patent and Trademark Office (USPTO). We then match the assignee's firm name to Compustat following the approach in \citet{kogan2017technological}. We use patent filings to measure a firm's patenting activities during a given quarter. Following \citet{skinner2023green}, we use International Patent Classification (IPC) codes to classify patents related to green technology. Similarly, by following \citet{toole2020inventing}, we identify IPC codes related to AI technology.\footnote{The IPC codes that we use to classify green and AI-related patents are available upon request.} As with the lobbying amounts, the distribution of patent filings is heavily skewed (the 90th percentile constitutes one filing per quarter). Therefore, we use an indicator variable $\textbf{1}\left(\text{\# Green Patent}>0\right)$ that takes the value of one when a firm files at least one green patent in the following quarter and zero otherwise.\footnote{For patent and lobby variables, unmatched firm-quarter observations are assumed to be zero. Nonetheless, we find that our results are robust to excluding unmatched observations or zero values (untabulated).} The variable $\textbf{1}\left(\text{\# AI Patent}>0\right)$ is defined similarly.

\section{Descriptive Statistics}
To provide preliminary evidence that GPT-based risk exposure measures capture risks in the intended ways, we start by examining descriptive statistics. We explore cross-industry variation in our risk scores and compare the time trends in our measures with those of bigram-based measures. We also investigate the face validity of our measures. 
\subsection{Descriptive Statistics and Industry-Level Averages}
In Table \hyperref[t1]{1}, Panel A, we report descriptive statistics for risk measures as well as for the variables used in our regression analyses. We observe that risk exposure \textit{assessments} are more than double in length compared to risk exposure \textit{summaries}. Consistent with \citet{hassan2019firm} and \citet{sautner2023firm}, for most firms, the levels of risk exposure are relatively close to zero. In particular, the median values of $PRiskAssess$, $CRiskAssess$, and $AIRiskAssess$ are 0.011, 0.001, and 0.000, respectively. However, there are also observations with large values. This result implies that in the majority of earnings calls, some risk information is disclosed even though it does not account for a large portion of the call. In addition, political risk is the most commonly mentioned risk, followed by climate risk. 

In Panel B, we present Pearson correlations among the risk measures, including the bigram-based political and climate risk exposures. As expected, bigram-based measures and GPT-based measures are positively correlated, although the correlation coefficients are relatively low (ranging from 0.116 to 0.147). Our corresponding assessment and summary measures are highly correlated with the correlation coefficients of 0.742 (political risk), 0.66 (climate risk), and 0.512 (AI risk). The political and climate risks also exhibit a correlation, with the correlation coefficients of 0.349 for summaries and 0.483 for assessments. These moderate correlations are expected because environmental and regulatory risks are inherently intertwined. The positive correlation is potentially further exacerbated by our sample period, during which a number of politically and environmentally significant events happen to coincide.

We verify that the political and environmental risk measures are not capturing the same content.\footnote{In Section 7.3, we conduct additional placebo tests and show that both measures capture a distinct risk dimension.} First, we estimate the correlation between the two assessment measures after demeaning them by the quarter average. The correlation decreases by a considerable amount from 0.483 (overall) to 0.351 (within a quarter). Second, we present word clouds in Figure \hyperref[fig5]{5}, which display very different keyword patterns for each risk type. Third, for non-zero risk exposure documents, we estimate the pairwise cosine similarity between the two risk assessments. Their average similarity is only 0.421.\footnote{As a benchmark, the cosine similarity between two continuous MD\&A reports is 0.8 to 0.9 \citep{brown2011large}. Also, for firms in the same industry, the pairwise similarity is, on average, 0.55 \citep{peterson2015earnings}.} Lastly, we manually check contents for a random sample of generated summaries and assessments and verify that the two risk categories deal with substantially different topics.

We further validate our measures by examining the two-digit SIC industries with the highest risk assessment scores (Figure \hyperref[fig3]{3}). For political risk, the high-risk industries include, for example, tobacco products and heavy construction. For climate risk, the high-risk sectors include coal mining, electricity and gas, textile, and paper products. In line with observed positive correlations between the two risk topics, there is some (expected) overlap in the high-risk industries (e.g., textile mill products, electricity \& gas). Lastly, for AI risk, the highest risk industry composition is quite different. Business services, engineering services, and electronic equipment are among the highest-ranked industries. We provide a heat map for one-digit SIC industry risk exposure in the Online Appendix, Figure 1. The results are consistent with those in Figure \hyperref[fig2]{2} and Figure \hyperref[fig3]{3}.

\subsection{Time Series Variation}
We next explore the time series properties of our risk measures, which are depicted in Figure \hyperref[fig4]{4}. Panel A presents political risk exposure based on GPT summaries and assessments, as well as a bigram-based measure based on \citet{hassan2019firm}. GPT-based and bigram-based measures exhibit similar time trends. Irrespective of the metric, political risk spikes in 2020, which is the year of the COVID-19 pandemic and the US Presidential Election. We also see another increase in political risk in 2022, during which the outbreak of Russia-Ukraine war took place. Thus, GPT-based measures appear to reflect notable political events in a timely and intuitive manner and are aligned with bigram-based measures.

Panel B of Figure \hyperref[fig4]{4} plots GPT-based and bigram-based climate risk exposure measures. All three climate risk measures co-move over time, although they are not as closely aligned as the political risk measures. We observe spikes in 2020, the first quarter of 2021, and the first quarter of 2022. The 2020 spike follows the UN Climate Action Summit,\footnote{During which climate activist Greta Thunberg delivered her widely resonating speech accusing world leaders of failing to tackle climate change.} and coincides with the outbreak of COVID-19 that introduced uncertainty regarding climate action. The first quarter of 2021 was impacted by the Texas power crisis, where a severe winter storm brought to light the vulnerability of the electricity supply to extreme weather events. Again, observable time trends in our GPT measures line up with notable environmental events.

Lastly, in Panel C, we plot AI-related risk summaries and assessments. Because no public bigram-based AI risk measure is available from prior studies, we do not have a natural benchmark. AI risk exposure exhibits an entirely different time trend than the two preceding risks. We observe a steady increase over time reflecting the heightened significance of AI technologies. A notable increase took place between the third quarter of 2019 and the second quarter of 2020. This period coincides with the release of Transformer-based BERT (Google) and GPT-2 (OpenAI) models, which have since been widely adopted by companies. The highest increase in AI risk occurred in the first quarter of 2023, immediately following the release of GPT3.5, which became viral around the world.\footnote{See \href{https://fortune.com/2023/03/01/a-i-earnings-calls-mentions-skyrocket-companies-say-search-cybersecurity-medicine-customer-service/}{this article} published in Fortune for anecdotal evidence.}

\subsection{Face Validity}

To examine the face validity of our measures, we read a number of generated summaries and evaluations for the three types of risk and confirm that they address the relevant topic. Appendix B presents several snippets from GPT-processed risk exposure documents. Each summary and assessment pair is from the same respective earnings call transcript. Assessments, as expected, contain richer contents that often include GPT's judgments based on the given context.

In example B1 (Appendix B), the political risk summary expresses a cautionary note that the recent government auction of HS1 project (high-speed railway line) in the UK creates exposure to political and regulatory uncertainties in Europe. In turn, the corresponding political risk assessment (see example B2) highlights that ``\textit{deficit reduction in these regions may lead to an increased flow of government disposals and potentially PFI (Private Finance Initiative) opportunities}." In the earnings call transcript, there is no mention of government disposals or PFI in conjunction with deficit reductions. Rather, PFI is mentioned in the later part of the executive's presentation. This example highlights GPT's ability to connect information pieces into a logical assessment.

The examples of climate risk summaries (B3) and assessments (B4) display a starker difference. The text of the climate risk summary only contains "NA," implying no direct mention of climate-related uncertainties. However, despite the lack of explicit discussion of climate risks, the corresponding climate risk assessment states that the company is subject to risks related to significant energy consumption and electronic waste. It also mentions a potentially high carbon footprint and the need for compliance with environmental regulations. 

Turning to the examples of AI Risk summaries (B5) and assessments (B6), we observe that the summary identifies the firm's active efforts to incorporate AI into its operations and provides more specific details. In contrast, risk assessment (B6) takes a more holistic view of whether the company's business is dependent on AI technologies.

In sum, the above-mentioned examples illustrate that the generated risk summaries and assessments are readable and generally contain relevant information when it comes to understanding firm-level risks. Further, risk summaries and assessments are distinct from each other in logically expected ways.

\subsection{Variance Decomposition}
In this subsection, we perform a variance decomposition of our proposed firm-level risk exposure measures into time, industry, and firm-level variation.\footnote{We use two-digit SIC industry classification. In untabulated analyses, we also use three-digit and one-digit level industry classifications and find consistent results.} The results are presented in Table \hyperref[t2]{2}, Panel A1, B1, and C1. The table shows the incremental R-squared values attributable to each set of fixed effects. While the descriptive evidence above reveals clear time trends and industry variation in our risk exposure measures, they jointly explain only a small portion of the variation in risk exposures. These results are closely in line with the findings in \cite{hassan2019firm}, suggesting that macroeconomic and sector-level risk variance is only the tip of the iceberg. There are some differences in the explained portion of the variance depending on the specific risk type and the measurement approach, but the general consistency is apparent and noteworthy. The results also imply that the variation attributable to firm-level factors is in the range of 80\% to 90\%, highlighting the importance of firm-level risk measurement.

We further decompose firm-level variance into time-varying and invariant portions. Specifically, we zoom in on residual variance by including firm-level fixed effects and report their corresponding R-squared values. Firm fixed effects explain between 21\% and 35\% of firm-level (residual) variance, thus implying that almost 65-79\% of variation is time-varying.\footnote{To further validate our measures, we estimate the measurement errors in our variables following \citet{sautner2023firm} (Online Appendix Table 1). Measurement errors for \textit{PRiskAssess} and \textit{CRiskAssess} are 2.71\% and 8.50\%, respectively. \textit{AIRiskAssess} has a relatively higher measurement error of 27.63\%, which is due to the high proportion of zero values in earlier years. Once we exclude zeros from analysis, the measurement error decreases to 6.59\%.} Pairwise comparisons of different types of risk measures (summary-, assessment-, and bigram-based) reveal that the assessment-based measures tend to be the more stable within a firm and across time.

In sum, although the risk measures exhibit notable aggregate time trends (as seen in Figure \hyperref[fig4]{4}), rich firm-level heterogeneity exists in exposures to these aggregate trends across different types of risks.

\section{Capital Market Consequences}
A valid measure of risk must exhibit association with volatility \citep{engle2004risk}. In this section, we examine associations between our risk proxies and two stock price volatility variables introduced previously, namely, implied volatility and abnormal volatility (described in Section 4.2). Specifically, we estimate the following OLS regression:
\begin{align}
    Volatility_{it+1} = \beta Risk_{it} + \gamma \textbf{X}_{it} + \delta_x + \varepsilon_{it},
\end{align}
where $Volatility_{it+1}$ is a proxy for the volatility of a firm $i$'s stock price in quarter $t+1$.\footnote{In the Online Appendix, we also use realized volatility, which is measured as the standard deviation of stock returns over a 90-day window, as an additional volatility variable.} $Risk_{it}$ is one of the firm-level risk exposure measures for firm $i$ in quarter $t$. Depending on the specification, we include a GPT summary-based risk exposure, \textit{RiskSum}, and a GPT assessment-based risk exposure, \textit{RiskAssess}, as well as a bigram-based measure, \textit{RiskBigram}. Prefix $P$, $C$, or $AI$ in front of $Risk$ denotes political, climate, and AI-related risks, respectively. $\textbf{X}_{it}$ is a vector of firm-quarter controls. Following \citet{hassan2019firm}, we control for the log of total assets in our main analysis. We also include an extensive set of additional controls in the Online Appendix and verify the robustness of our results. $\delta_x$ denotes various fixed effects. Specifically, we estimate two different sets of fixed effects in our main tests: (1) time ($\delta_q$) and industry ($\delta_s$) fixed effects, and (2) the interactions of time and industry fixed effects ($\delta_q \times \delta_s$). We do not control for firm-fixed effects, as firm-level variation in our measures is of interest. However, in the Online Appendix, we confirm that our results are robust to controlling for firm fixed effects. Robust standard errors are clustered at the firm level. All continuous variables are winsorized at 1\% and 99\% levels.

\subsection{Political Risks}

We start discussing our main results with respect to political and climate risk because they received considerable attention in prior studies (and because bigram-based benchmarks are available for comparison purposes). Thus, we confine our main sample to the 2018-2021 period as the bigram-based measures are updated until early 2022 \citep{hassan2019firm, sautner2023firm}.

The estimates for political risk are reported in Table \hyperref[t3]{3}. As shown in columns (1) and (2), both GPT-based measures of political risk, \textit{PRiskSum} and \textit{PRiskAssess}, exhibit positive and highly statistically significant associations with both implied volatility and abnormal volatility. This result holds with industry fixed effects (Panel A) and with industry-time fixed effects (Panel B). In columns (3) and (4), we add the bigram-based risk measure as an additional control variable to the regressions.\footnote{This measure is statistically significant when used on its own.} We find that the effect of our GPT-based measures does not attenuate and that GPT-based measures subsume information in \textit{PRiskBigram}. Overall, this evidence suggests that LLM-based risk measures are effective at capturing political risk and that they dominate dictionary-based measures.

We next turn to the comparison of GPT-based summaries vs. assessments in explaining volatility. When adding the risk measures simultaneously to the same regression, as we do in column (5), we find that the risk assessment-based measure dominates the summary-based measure for both implied and abnormal volatility. This result is important as it implies that GPT's ability to synthesize general knowledge acquired from its training is valuable when assessing political risk exposure.\footnote{For abnormal volatility, we observe negative and significant coefficient for \textit{PRiskSum} when including industry-time fixed effects. This phenomenon arises due to the high positive correlation between \textit{PRiskAssess} and \textit{PRiskSum}. The effect is expected when both proxies contain correlated measurement errors and one of the proxies is a dominant one.} Overall, these findings reinforce the notion that LLMs can produce insights by synthesizing the input text and general knowledge. 

In terms of the economic significance of the results, a one standard deviation increase in \textit{PRiskAssess} translates to 0.04 standard deviations increase in implied volatility (using column (5) of Panel A), which is a non-trivial amount. Similarly, one standard deviation increase in \textit{PRiskAssess} is associated with 0.06 standard deviations increase in abnormal volatility (using column (5) of Panel A).

\subsection{Climate Risks}
Next, we estimate the same model using the climate risk proxies as our independent variables. The results are reported in Table \hyperref[t4]{4}. GPT-based measures bear a positive and statistically significant association with stock price volatility both on a standalone basis or when used jointly with \textit{CRiskBigram}. The statistical power to detect climate risk is highest in the case of GPT-based assessments, \textit{CRiskAssess}, which exhibit several times higher t-values compared to the summary-based measure \textit{CRiskSum}. Furthermore, when the three proxies are included in the model simultaneously, \textit{CRiskAssess} is the only one that consistently remains statistically significant and exhibits the expected sign. These results are consistent regardless of the fixed effect structures and hold for both implied volatility and abnormal volatility. 

Overall, LLMs perform well at identifying climate-related concerns and subsume bigram-based measures. Furthermore, the analysis also confirms the value of LLM's general knowledge -- GPT assessments -- in evaluating climate risks, going beyond the information discussed during the conference calls. 

\subsection{AI-Related Risks}

Lastly, Table \hyperref[t5]{5} examines the association between AI-related risk exposure and stock price volatility. As there is no established bigram-based AI risk measure in the literature, we only analyze the GPT-based measures. In our primary sample, we find limited evidence that AI-related risk is associated with stock price volatility. For example, we do not find a statistically significant relation between AI risk proxies and implied volatility. In the case of abnormal volatility, we only find significant associations in the case of \textit{AIRiskSum}. This result is not surprising as AI risk has emerged as an economically important phenomenon only very recently. Indeed, from 2018 to 2021, nearly 60\% of the observations have zero AI-related risk exposure. In the following section, we explore whether AI risk becomes significant in explaining stock price volatility in the most recent period.

\subsection{Analysis Outside GPT's Training Period}

We note that all of the analyses in our study are effectively out-of-sample as the language model is not trained to predict any of the outcome variables that we examine. Nevertheless, one potential concern is that our main sample overlaps with the GPT model's training period, which extends to September 2021. It is possible (though not known) that GPT had seen the transcripts of earnings calls during its training, which may give the model an edge in generating risk summaries and assessments. 

To address this concern, we conduct ``true" out-of-sample tests. In particular, we use a sample of conference calls from January 2022 to March 2023, which is outside of the GPT's training sample. We present the results of this analysis in Table \hyperref[t6]{6}. Panels A and B use GPT-based political risk and climate risk proxies, respectively, as their main independent variables. Even in this more limited sample, we continue to find positive and highly significant associations between GPT-based risk measures and firms' stock price volatility. The only exception is that \textit{PRiskSum} and \textit{CRiskSum} lose statistical significance when examining abnormal volatility (column (3) of both panels). Nevertheless, they remain positive and retain their economic magnitudes. Most importantly, the assessment-based exposure measures are significantly associated with both volatility across all model specifications.

For AI Risk, reported in Panel C, in fact, we find stronger results during this more recent time window. Specifically, we observe that both \textit{AIRiskSum} and \textit{AIRiskAssess} become positive and significantly significant determinants of volatility. This finding directly maps into our evidence in Figure \hyperref[fig4]{4C}, which indicates that AI risk has soared in recent quarters. In terms of economic magnitudes, a one standard deviation increase in \textit{AIRiskAssess} is associated with 0.03 standard deviations increase in implied volatility and 0.03 standard deviations increase in abnormal volatility, which is once again a sizable effect.

These findings lead to two important insights. First, GPT's ability to produce valid risk exposure measures cannot be attributed to its in-sample knowledge. Rather, it suggests that large language models are useful for investors' decision-making process when used outside of the training window. Second, the fact that AI risk is correlated with volatility only in the period after it soars constitutes further evidence of the construct validity of our measures. 

\subsection{Robustness checks}

We also conduct a battery of robustness tests in the Online Appendix. First, motivated by our findings in Section 5.4, we include \textit{firm} fixed effects into the model (Online Appendix Table 2). Second, we expand the set of our control variables to include leverage, cash holdings, tangible assets, profitability, and capital expenditures (Online Appendix Table 3). In both sets of robustness tests, we continue to find strong results for the GPT-based assessment measures, which outperform the dictionary-based proxies. Third, we use realized volatility (instead of implied volatility) as another dependent variable. We show that \textit{PRiskAssess} and \textit{CRiskAssess} are positively associated with this volatility measure (Online Appendix Table 4). Fourth, to mitigate the concern that extreme risk exposure values are driving our results, we plot the association between the risk assessment variables and implied volatility in Online Appendix Figure 2. We randomly sample observations whose risk assessment values are within 10\% and 90\% percentile and show that the positive association remains.

Overall, our analysis lends support to the notion that GPT-based firm-level political risk exposure and climate change risk exposure are useful in capturing corporate risks. These measures are superior to traditional bigram-based measures, indicating that GPT can produce substantially informative risk assessments. 

\section{Risk Exposures and Firm Decisions}

An alternate approach to evaluate the economic usefulness of GPT-based risk measures involves examining their ability to explain firms' economic decisions. We focus on capital investment decisions, followed by the examination of corporate actions aimed at diminishing risk exposures, such as lobbying or developing intellectual property.

\subsection{Investment decisions}
\label{sec:investment}
Theoretical considerations suggest that heightened uncertainty regarding a firm's future commands a risk premium that investors require in exchange for capital \citep[e.g.,][]{pastor2013political}. Consequently, higher risk exposure should make investments in capital projects harder to finance. Supporting this notion, \citet{baker2016measuring} and \cite{hassan2019firm} show a decline in capital expenditures in response to escalating political uncertainty.\footnote{In theory, investments are also expected to decline in response to increased uncertainty because the option to wait becomes more valuable \citep{dixit1994investment, bloom2007uncertainty}. Although the cost of capital effect on investment applies to all risk types, there is a countervailing effect when capital expenditures become an important part of a firm's strategy to counteract a specific risk. For example, companies might be less inclined to rely on tangible investments to alleviate political risk, favoring methods like lobbying or political donations instead. However, to address risks posed by AI, significant infrastructure investments are likely to be needed to bolster resilience to AI risk.}

To examine the association between risk exposures and investment, we estimate the following OLS regression:
\begin{align}
    Investment_{it} = \beta Risk_{it} + \gamma \textbf{X}_{it} + \delta + \varepsilon
\end{align}
where $Investment_{it}$ is capital expenditure intensity (see Section 4.3 for more details). We include quarter and industry fixed effects. The remaining variables are the same as in Equation (3). We winsorize continuous variables at 1\% and 99\%. Standard errors are clustered at the firm level.
 
Table \hyperref[t7]{7} presents the results. In Panel A, columns (1) and (2) show that \textit{PRiskSum} and \textit{PRiskAssess} have a negative and statistically significant association with capital expenditures. The coefficient of \textit{PRiskAssess} is not only larger than that of \textit{PRiskSum} but also more significant. When we jointly include GPT-based and bigram-based measures in columns (3) and (4), the GPT-based measures remain significantly associated with investments and bigram-based measures are not. In line with prior evidence, we also find that \textit{PRiskAssess} is the dominant proxy, as can be seen from column (5). In terms of economic significance, a one standard deviation increase in \textit{PRiskAssess} translates to a 1.3\% decrease in investment, on average (based on estimates in column (5)).

In Panel B, we present results for climate risk and observe a similar pattern as in Panel A. \textit{CRiskAssess} bears a negative relation with investments, which remains significant with and without the bigram-based measure or \textit{CRiskSum} in the same regression. In terms of economic magnitude, a one standard deviation increase in \textit{CRiskAssess} is associated with a 0.8\% decrease in average investment.

In Panel C, we do not find a significant relation between AI risk and investments. In both columns, we find positive yet not statistically significant coefficients on both risk exposure measures. 

To shed more light on the relationship between AI-related risk and capital expenditure, we examine the most recent period starting in January 2022 and ending in March 2023, during which AI risks skyrocketed, and we perform several cross-sectional tests. In particular, we examine whether companies that are financially constrained, i.e., lack internal funds to finance investments, have a decrease in their capital expenditures in response to increasing AI risks (cost of capital channels). In contrast, financially unconstrained firms are expected to increase their investment to tackle the AI challenges.

We partition the sample based on the levels of cash holdings and leverage measured as of 2021 and report the analysis in the Online Appendix, Table 11. We observe that the least financially constrained firms exhibit the largest increase in capital expenditures in response to \textit{AIRiskAssess}. In contrast, we find that the most constrained firms reduce their investments. These findings help to understand the seemingly counterintuitive impact of AI risk on investments. 

Overall, our results indicate that GPT-based risk exposure measures are useful in explaining corporate investment decisions and do so in line with theoretical priors. However, the effect on investments is nuanced and depends on the type of risk.

\subsection{Firm Responses to Mitigate Risk Exposures} 

Firms are expected to proactively mitigate risks by implementing risk-specific countermeasures. In this subsection, we turn to the examination of firms' actions to mitigate their risk exposures. To do so, we focus on three outcome variables that correspond to three different risk types. First, we gauge our political risk measures by probing their ability to explain companies' lobbying activities \citep{peltzman1976toward}. Second, we investigate firms' reactions to climate threats by examining the issuance of green patents \citep{blyth2007investment, stern2021innovation}. Finally, we measure firms' responses to AI risk by focusing on AI-related innovation (the filing of AI-related patents).

We estimate the following OLS regression:
\begin{align}
    Response_{it+1} = \beta Risk_{it} + \gamma \textbf{X}_{it} + \delta + \varepsilon
\end{align}
where $Response$ is one of the three indicator variables discussed in Section 4.3 in more detail: $\textbf{1}(\text{\$ Lobby Amount}>0)$ is an indicator variable that equals one if firm $i$ lobbying expenditures are greater than zero in the quarter $t+1$ and zero otherwise, $\textbf{1}(\text{\# Green Patent} >0)$ is an indicator variable that equals one when firm $i$ files at least one green patent in the quarter $t+1$ and zero otherwise, and $\textbf{1}(\text{\# AI Patent} >0)$ is an indicator that equals one when firm $i$ files at least one AI-related patent in the quarter $t+1$ and zero otherwise.
We expect a positive association between these variables and their corresponding risk exposure measures. As before, we use quarter and industry fixed effects and the same set of control variables as in Equation (3). 

Table \hyperref[t8]{8} presents how firms respond to a change in firm-level risk exposure. Panel A focuses on lobbying activity in response to political risk. We find a positive and significant association between \textit{PRiskAssess} and lobbying, whereas \textit{PRiskSum} only shows a limited association. When \textit{PRiskAssess} is included jointly with \textit{PRiskBigram}, the two measures both remain incrementally informative. A one standard deviation increase in \textit{PRiskAssess} is associated with a 1.40\% point increase in the likelihood of lobbying in the following quarter (based on estimates in column (5)).\footnote{Note that \textit{PRiskSum} is negative and statistically significant in column (5). To the extent the same information is repeated in the assessment summaries, the model performs better by filtering it out from the assessment-based measure.}

Panel B studies the association between green patents and climate risk exposure. In column (2), we find a positive and significant association between \textit{CRiskAssess} and green patent filings. When we include all three risk proxies in one regression, \textit{CRiskAssess} comes out as a dominant proxy. In terms of economic magnitude, based on our estimates in column (5), a one standard deviation increase in \textit{CRiskAssess} is associated with a 0.73\% point increase in the likelihood of filing a green patent in the following quarter.

Finally, in Panel C, we repeat the analysis for AI risk and AI patent filings. We find that one standard deviation increase in \textit{AIRiskAssess} is associated with a 2.39\% point increase in the likelihood of filing at least one AI patent in the subsequent quarter.

Overall, we find that companies respond to risks captured by our measures by taking actions to mitigate them. We view this finding as further evidence of the validity and economic usefulness of GPT-based risk exposures. 

\subsection{Further Out-of-Sample Analysis}

Similar to Section 6.4, we perform a subsample analysis from January 2022 to March 2023 to ensure that our results are not attributable to GPT seeing the underlying data during its training phase. Table \hyperref[t9]{9} presents the results of this analysis. Overall, our results are consistent with those in Tables \hyperref[t7]{7} and \hyperref[t8]{8}, addressing concerns about possible in-sample bias. In fact, for lobbying activity, \textit{PRiskSum} becomes even more significant in the out-of-sample test.

One notable difference in this analysis is that AI risk, on average, shows a positive and significant association with investments in the most recent year. This finding reconciles with the additional analysis in section \ref{sec:investment} and is noteworthy because standard theory would predict that their investments should decrease with risks. As discussed previously, the likely explanation for this finding is that companies respond to soaring AI risks by making investments in AI-related technology and infrastructure. 

\subsection{Robustness Checks} 

We perform several robustness checks. We gauge the robustness of our results by including firm fixed effects (Online Appendix Table 6), as well as by including an extensive set of controls (Online Appendix Table 7).\footnote{In Tables \hyperref[t7]{7} and \hyperref[t8]{8}, we only report the results using industry and quarter-fixed effects. However, the results are almost identical when we allow for the interactions of quarter and industry fixed effects.} Our results are similar in these alternative specifications. To mitigate the concern that extreme values are driving our results, we visualize the linear association between risk exposure measures and investments after excluding extreme values (Online Appendix Figure 3).

To bolster our findings that each risk is exclusively associated with its corresponding corporate action (and not just proxies for general risk), we perform several placebo tests. Specifically, we regress a lobbying activity indicator on \textit{CRiskAssess} and \textit{AIRiskAssess}, a green patent indicator on \textit{PRiskAssess} and \textit{AIRiskAssess}, and an AI patent indicator on \textit{PRiskAssess} and \textit{CRiskAssess}. We report the results in Online Appendix Table 9. In general, each economic outcome is most significantly associated with its corresponding risk measure. One exception is \textit{AIRiskAssess}, which is also positively associated with lobbying and green patent activities. 

Finally, because patenting and lobbying activity variables are heavily skewed, we use their natural log as dependent variables. Additionally, following \citet{COHN2022529}, we also use Poisson regressions for patent counts (Online Appendix Table 10). Our results remain qualitatively similar. Overall, our findings highlight that firms respond to risk by subsequently mitigating their risk exposure.



\section{Relative Importance of Risks Over Time}
In this section, we explore the relative importance of different types of risk over time. To do so, we use the four-quarter rolling window to estimate the following model:\footnote{Our first estimation period starts in the first quarter of 2018 and ends in the fourth quarter of 2018, and so on. The last estimation period ends in the first quarter of 2023.} 
\begin{align}
    Implied\_Volatility_{it} = & \beta_{1t} PRiskAssess_{it} + \beta_{2t} CRiskAssess_{it} + \beta_{3t} AIRiskAssess_{it} \nonumber \\
    & + \gamma_t \textbf{X}_{it} + \delta_q + \delta_s + \varepsilon_{it} 
\end{align}

where the risk proxies are based on risk assessments and are included either simultaneously or one at a time; $\delta_q$ is quarter fixed effect and $\delta_s$ is industry fixed effect. For each rolling regression, we report coefficient $\beta_t$ and the corresponding \textit{t}-values. 

Table \hyperref[t10]{10} presents the estimates. In Panel A, we estimate the importance of each risk type on a stand-alone basis, i.e., include the risk proxies one at a time. We also visualize the time-series variation in \textit{t}-values in Figure \hyperref[fig6]{6A}. In Panel B, however, we compare the relative importance of each risk type after controlling for the effect of the other types, i.e., include the three risk proxies simultaneously. We also visualize the \textit{t}-value time trends in Figure \hyperref[fig5]{6B}.

Overall, the two figures display similar patterns. Consistent with Figure \hyperref[fig4]{4}, we observe a clear upward trend in \textit{t}-values associated with \textit{AIRiskAssess}. \textit{AIRiskAssess} is insignificant in both panels based on a 10\% significance level until the second quarter of 2022. However, in the last four rolling windows, \textit{AIRiskAssess} becomes increasingly important. We also show that climate change risk and political risk are both highly significant during 2020. However, in both panels, climate change risk exhibits higher statistical significance than political risk in 2020. This trend reverses in 2021, during which political risk becomes relatively more significant than climate change risks. In sum, while AI risk is a clearly emerging force, the other types of risks exhibit high and low periods throughout our sample period.

\section{Equity Market Pricing}
In our last set of tests, we probe the asset pricing implications of our GPT-based risk exposure measures. In theory, higher risk exposures should be associated with higher expected equity returns. Establishing this link is challenging in our setting as the asset pricing methodology requires a relatively long time series. We use our entire sample period, starting in January 2018 and ending in March 2023, for this test. As this period is still relatively short, our results (t-statistics) are likely to understate the significance of risks picked up by our proxies. 

We test the pricing of GPT-based proxies by running \citet{fama1973risk} regressions and performing portfolio analysis for our risk variables. In this analysis, we focus on risk assessment-based measures as our prior tests imply their dominance. As \citet{fama1973risk} regressions typically use annual characteristics, we annualize our risk proxies by taking their average across four quarters each year ($PRiskAssess^{ann}$, $CRiskAssess^{ann}$, and $AIRiskAssess^{ann}$).\footnote{We exclude observations with zero annualized risk exposures since they are likely not to feature any discussion of political, climate change, or AI-related risks during earnings calls (or, alternatively, they might be instances where the model fails to generate meaningful output).} We construct our portfolios on March 31 of the subsequent year to allow for three or more months for stock prices to incorporate information disclosed during earnings calls (e.g., we compute annualized risk exposure over 2021 and form the portfolios on March 31, 2022). 

We regress monthly stock returns on the natural log of each risk exposure, log of the market value of equity (log(\textit{ME})), log of book-to-market ratio (log(\textit{BE/ME})), operating profitability (\textit{Profitability}), investment (\textit{Investment}), lagged one-month return ($r_{0,1}$), and lagged annual return after excluding the most recent month ($r_{2,12}$) \citep{fama2015five}.\footnote{\textit{Investment} is given by growth in total assets; \textit{Profitability} is (total revenue – cost of goods sold – (sales, administrative expense – R\&D expense)) scaled by total assets; log(\textit{ME}) is the natural logarithm of the market value, and log(\textit{BE/ME}) is the natural logarithm of the book-to-market ratio.} We report Newey-West \textit{t}-values with a lag of 3. All continuous independent variables are trimmed at 1\% and 99\% levels.

We report \cite{fama1973risk} analysis in Table \hyperref[t11]{11}, Panel A. After controlling for asset characteristics, all three risk exposure proxies exhibit positive coefficients. The coefficients on $CRiskAssess^{ann}$ (0.211) and $AIRiskAssess^{ann}$ (0.317) are positive and statistically significant at conventional levels. The coefficient on $PRiskAssess^{ann}$ is positive (0.077), yet not statistically significant (\textit{t}-value = 1.36). Overall, the positive associations support the pricing of the corresponding risks in equity markets.

For the portfolio analysis, we construct quintile portfolios based on the values of each annualized firm-level risk exposure measure. As earlier, we form portfolios on March 31 of the subsequent year relative to risk exposure and hold them for one year. 

Table \hyperref[t11]{11}, Panel B presents equal-weighted high-minus-low portfolio alphas based on \citet{fama2015five}’s five-factor model. Columns (1), (2), and (3) report monthly alphas from $PRiskAssess^{ann}$, $CRiskAssess^{ann}$, and $AIRiskAssess^{ann}$, respectively. We observe that portfolio alphas almost monotonically increase across risk exposure quintiles. Accordingly, all three columns report positive high-minus-low alphas. For political risk exposure, the annualized alpha is 5.28\%, which is economically sizable yet does not attain statistical insignificance (t-value = 1.51). For climate change risk exposure, the annualized alpha is 6.72\% and is statistically significant (t-value = 1.90). For AI-related risk exposure, the annualized alpha is 6.36\% and is also statistically significant (t-value = 2.31).

Overall, our results suggest that GPT-based risk exposure measures successfully extract firm-level risks that are priced in equity markets. Further, we show that such risk exposure measures are associated with sizable alphas unexplained by the five-factor asset pricing model.

\section{Conclusion}
In this paper, we evaluate whether recent advances in AI technology can help investors assess critical aspects of corporate risks. We evaluate these risks based on information disclosed during companies' earnings calls. More specifically, we use a generative Large Language Model, GPT 3.5 Turbo, to develop and validate three proxies for firm-level exposure to political, climate, and AI-related risks, all of which have been of primary concern to firms' stakeholders in recent years. We also investigate LLM's ability to leverage general knowledge to enhance summaries of risk-related content from conference calls by adding their own insights. 

Our measures of firm-level risk exposures exhibit large within-firm variation and move across industries and over time in intuitive ways. More importantly, each of the three measures is a powerful predictor of future stock price volatility and helps in explaining firms' policies, such as investment and innovation. Furthermore, GPT-based assessments are consistently more informative about firms' risks as compared to GPT summaries, highlighting the value of LLM's general knowledge. We also find that GPT-based risk measures consistently subsume the existing bigram-based risk proxies when subjected to the task of predicting stock market volatility and explaining economic decisions. Our findings are robust to a number of research design choices and hold outside of the training window of the GPT model.

We conclude that generative AI technology enables users to obtain valuable insights about firm-level risks at a relatively low cost. The generated risk summaries and evaluations are not only human-readable but also easily convertible to risk exposure scores. Overall, our study highlights the economic usefulness of generative AI tools, such as ChatGPT, in measuring in providing useful input into investors' decisions based on unstructured textual information \citep{kim2023bloated}.

We also recognize limitations that come with the reliance on generative LLMs. GPT outcomes can be sensitive to prompt quality, and prompt engineering becomes an important starting point. Unless provided with relevant context, GPT may generate incorrect evaluations or provide random answers (known as hallucinations).\footnote{Even with perfect prompts, GPT might still generate incorrect information. This is because GPT is a language model whose primary purpose is completing a sentence in the most natural manner. This process is inherently \textit{not} deterministic, and GPT might produce incorrect answers.} We recommended that users compare and reconcile GPT's answers with the corresponding source documents when making decisions. Nevertheless, in a large sample, the usefulness of generative AI in assessing risks is apparent.

\newpage
\onehalfspacing
\setlength\bibsep{0pt}
\bibliographystyle{elsarticle-harv}
\bibliography{custom}
\newpage

\section*{Appendix A. Example of SK Telecom}\phantomsection\label{a1}
The following is an excerpt from the GPT-processed political risk exposure assessment from SK Telecom's 2018 Q1 earnings call transcript.\\

The firm is subject to political and regulatory risks and uncertainties related to its 5G CapEx plans and the discussions within the National Assembly regarding the separation of handset contracts and telecom service contracts.

[…] Regarding the discussions on the separation of handset contracts and telecom service contracts, the company states that it is difficult to predict whether it will be introduced and the impact it may have. However, the company believes that if introduced, it could resolve issues related to the blending of service fees and handset pricing. The company also expects a shift in the market competition paradigm towards prices and service quality, leading to greater user benefits. The company emphasizes the need for measures that minimize negative effects on the overall ecosystem, including the distribution structure. The company looks forward to a reasonable conclusion being generated through the discussion and reviewing process, with a focus on market competition and enhancing customer convenience.

[…] The analyst mentioned the government's talk about the separation of device contracts and service contracts, which could potentially affect the company's business model. However, the executives mentioned that there are still ongoing discussions and it is uncertain what will happen. The company has its own stance on the issue but is open to various options before making a decision. This indicates that the company may be affected by new regulations in the future, but the extent of the impact is uncertain at this point.

\newpage 

\section*{Appendix B. Sample Snippets}\phantomsection\label{a2}
\subsection*{B1. Political Risk Summary}
The company is subject to political and regulatory risks and uncertainties in Europe and North America. The recent government auction of HS1 in the U.K. is mentioned as an example.

\subsection*{B2. Political Risk Assessment}
The firm is subject to political and regulatory risks and uncertainties in Europe and North America. The focus on deficit reduction in these regions may lead to an increased flow of government disposals and potentially PFI (Private Finance Initiative) opportunities. The recent government auction of HS1 in the U.K. is mentioned as an example. Additionally, the flow of non-core disposals by corporate and financial institutions is continuing, as evidenced by the firm's recent investment in Eversholt, which was purchased from HSBC. These political and regulatory factors could impact the firm's operations and investment opportunities in these regions. 

\subsection*{B3. Climate Change Risk Summary}
NA

\subsection*{B4. Climate Change Risk Assessment}
It is worth noting that the company's use of leading-edge technologies in wafers, silicon wafers, substrates, and packaging may have implications for their environmental footprint. These technologies often require significant energy and resource consumption during production and may generate electronic waste at the end of their lifecycle. Additionally, as the company's networking products are being used in the creation of Ethernet fabric for AI clusters, there may be indirect environmental risks associated with the energy consumption and carbon footprint of these clusters. It is important for the company to consider the sustainability of their networking products and ensure they are aligned with environmental regulations and standards.

\subsection*{B5. AI-Related Risk Summary}
Based on the given information, the firm is heavily dependent on AI technologies and is actively incorporating AI into every layer of its stack, including productivity and consumer services. The executives mention that they believe the next big platform wave is AI and that they are working on building training supercomputers and inference infrastructure. They also mention specific AI capabilities in their products, such as robotic process automation and workflow automation, as well as the incorporation of AI in their consumer services. The firm has a partnership with OpenAI and is excited about their innovation and commercialization of products. 
\subsection*{B6. AI-Related Risk Assessment}
Based on the given information, it is clear that Microsoft is heavily invested in emerging technologies, particularly artificial intelligence (AI). The company highlights its leadership in the AI era and its commitment to developing AI-powered products and services. Microsoft's Azure platform is being used by customers and partners to train state-of-the-art AI models and services, and the company is positioning itself as a leader in AI with its powerful AI supercomputing infrastructure. Additionally, Microsoft's AI services, such as Azure ML, have seen significant revenue growth, indicating a strong demand for AI capabilities. The company is also leveraging AI in its developer tools, such as GitHub Copilot, which is an AI-powered product that transforms developer productivity. Furthermore, Microsoft is integrating AI into its business applications, such as Dynamics 365, to help businesses digitize their operations and improve efficiency. Overall, Microsoft's business is heavily dependent on AI technologies, and the company is at the forefront of AI innovation. 
\end{sloppypar}

\newpage
\onehalfspacing
\section*{Figure 1. Time Trend in SK Telecom's Political Risk Exposure}\phantomsection\label{fig1}
\begin{spacing}{0.9}
{\footnotesize\noindent This figure shows the time trend in SK Telecom's political risk exposure from 2018 to 2022. The solid line represents GPT-based risk exposure assessment (\textit{PRiskAssess}) and the dotted line represents the bigram-based risk exposure score by \citet{hassan2019firm} (\textit{PRiskBigram}).}
\end{spacing}

\begin{figure}[hbt!]
\centering
\captionsetup{labelformat=empty}
\includegraphics[width=130mm]{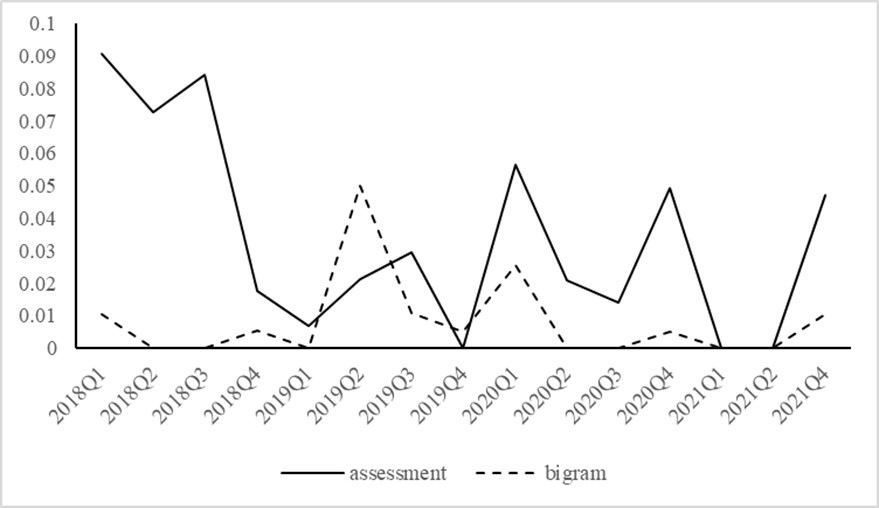}
\caption{Figure 1. Time Trend in SK Telecom's Political Risk Exposure}
\end{figure}

\onehalfspacing
\section*{Figure 2. Measuring Risks with Generative AI}\phantomsection\label{fig2}
\begin{spacing}{0.9}
{\footnotesize\noindent This figure summarizes how we process earnings call transcripts with GPT to generate firm-level exposure measures. Refer to Section 3 for detailed explanation.}
\end{spacing}

\begin{figure}[hbt!]
\centering
\captionsetup{labelformat=empty}
\includegraphics[width=130mm]{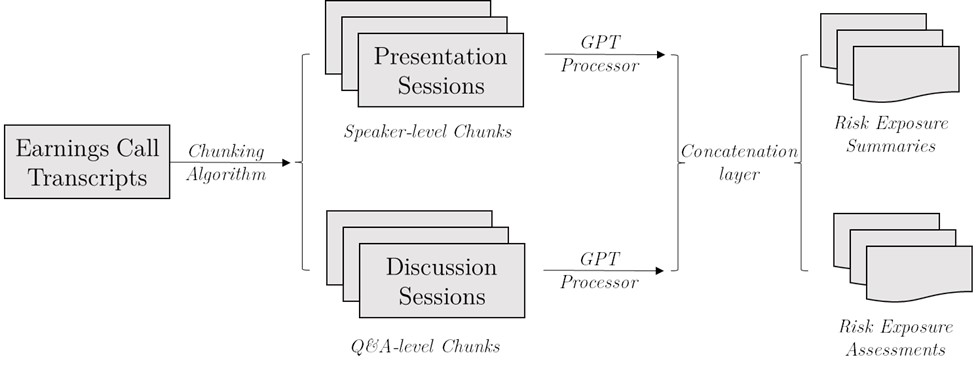}
\caption{Figure 2. Measuring Risks with Generative AI}
\end{figure}
\clearpage
\newpage

\onehalfspacing
\section*{Figure 3. Industry Averages of Risk Exposure Assessments}\phantomsection\label{fig3}
\begin{spacing}{0.9}
{\footnotesize\noindent This figure shows the SIC two-digit level industry averages of GPT-based risk exposure assessments (RiskAssess). We regress \textit{RiskAssess} on dummy variables that represent each industry and report eight industries with the largest coefficients. We also plot 95\% confidence intervals. Standard errors are clustered at the firm-level. 3A shows industry-level averages of political risk exposure, 3B shows climate change risk exposure, and 3C shows AI-related risk exposure.}
\end{spacing}

\begin{figure}[hbt!]
\centering
\captionsetup{labelformat=empty}
\includegraphics[width=110mm]{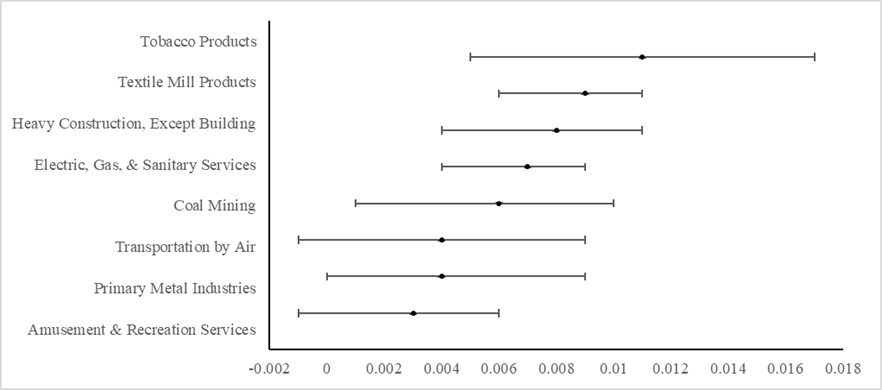}
\caption{Figure 3A. Industry Averages of \textit{PRiskAssess}}
\end{figure}

\begin{figure}[hbt!]
\centering
\captionsetup{labelformat=empty}
\includegraphics[width=110mm]{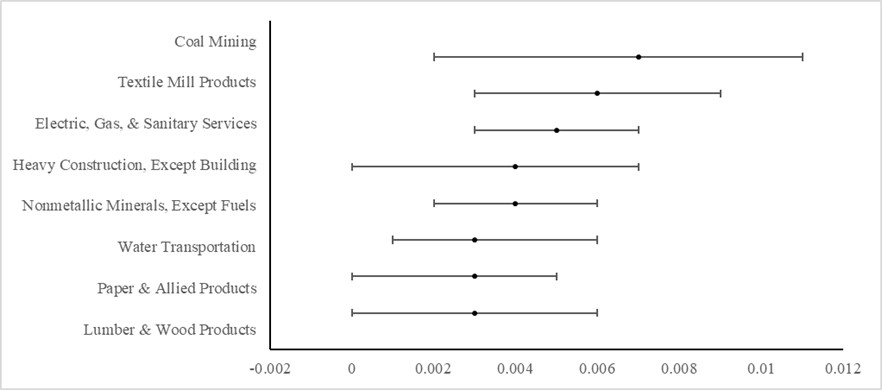}
\caption{Figure 3B. Industry Averages of \textit{CRiskAssess}}
\end{figure}

\begin{figure}[hbt!]
\centering
\captionsetup{labelformat=empty}
\includegraphics[width=110mm]{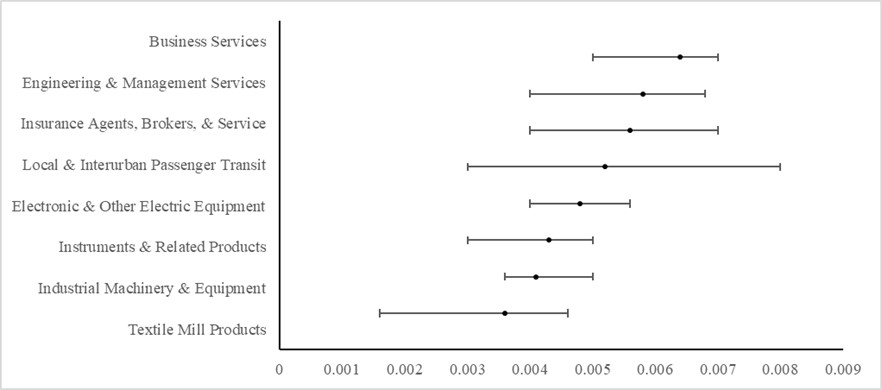}
\caption{Figure 3C. Industry Averages of \textit{AIRiskAssess}}
\end{figure}

\clearpage
\newpage

\onehalfspacing
\section*{Figure 4. Time Trend of Risk Measures}\phantomsection\label{fig4}
\begin{spacing}{0.9}
{\footnotesize\noindent This figure shows the time series variation in firm-level risk exposure measures. 4A shows political risk exposure measures. 4B shows climate change risk exposure measures. We include the bigram measure, GPT-based summary measure, and GPT-based assessment measure. 4C shows AI-related risk exposure measures. For 4C only, we include GPT-based summary measure and GPT-based assessment measure. Shaded areas denote notable economy-wide events related to each risk.}
\end{spacing}

\begin{figure}[hbt!]
\centering
\captionsetup{labelformat=empty}
\includegraphics[width=100mm]{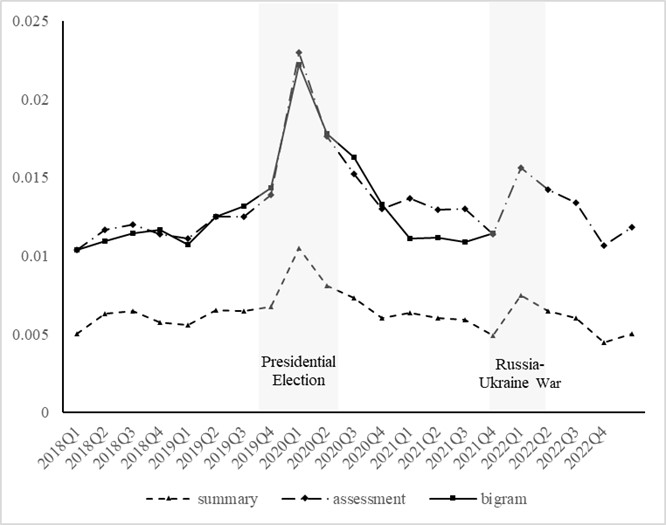}
\caption{Figure 4A. Time Trend of \textit{PRiskAssess}}
\end{figure}

\begin{figure}[hbt!]
\centering
\captionsetup{labelformat=empty}
\includegraphics[width=100mm]{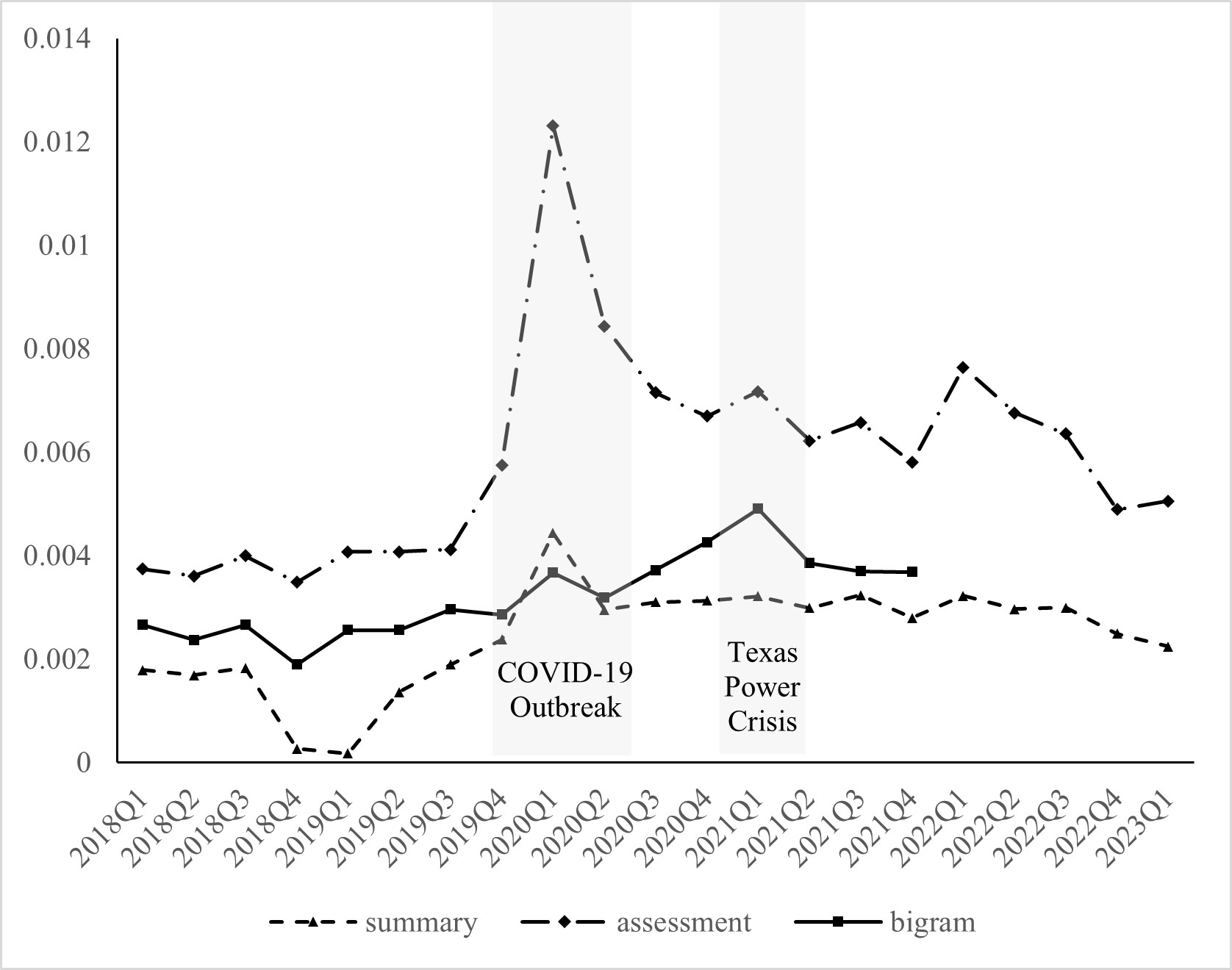}
\caption{Figure 4B. Time Trend of \textit{CRiskAssess}}
\end{figure}

\begin{figure}[hbt]
\centering
\captionsetup{labelformat=empty}
\includegraphics[width=100mm]{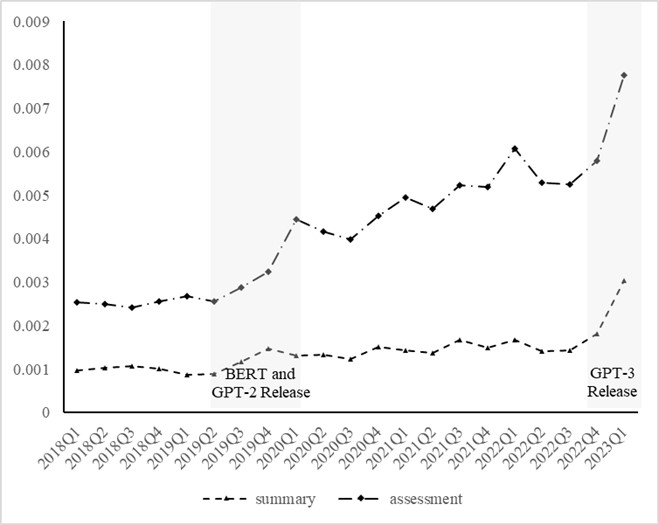}
\caption{Figure 4C. Time Trend of \textit{AIRiskAssess}}
\end{figure}

\clearpage
\newpage

\onehalfspacing
\section*{Figure 5. Word Clouds}\phantomsection\label{fig5}
\begin{spacing}{0.9}
{\footnotesize\noindent This figure shows the word clouds extracted from the underlying documents of $PRiskAssess$, $CRiskAssess$, and $AIRiskAssess$.}
\end{spacing}

\begin{figure}[hbt!]
\centering
\captionsetup{labelformat=empty}
\includegraphics[width=85mm]{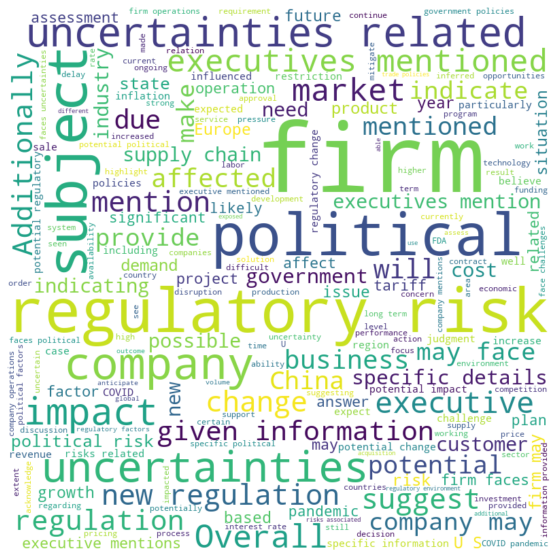}
\caption{Figure 5A. Word Cloud of \textit{PRiskAssess}}
\end{figure}

\begin{figure}[hbt!]
\centering
\captionsetup{labelformat=empty}
\includegraphics[width=85mm]{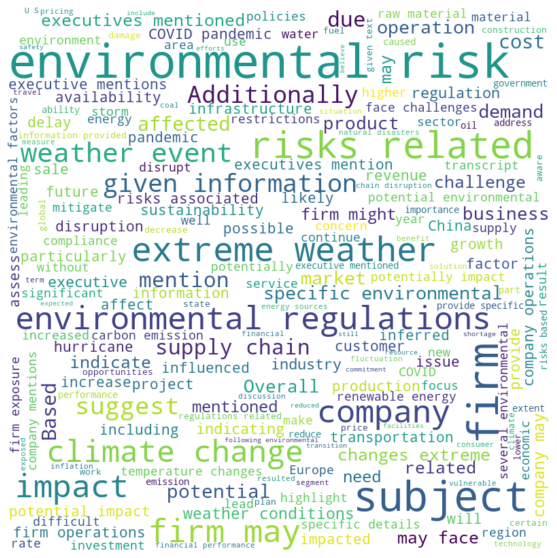}
\caption{Figure 5B. Word Cloud of \textit{CRiskAssess}}
\end{figure}

\begin{figure}[hbt]
\centering
\captionsetup{labelformat=empty}
\includegraphics[width=85mm]{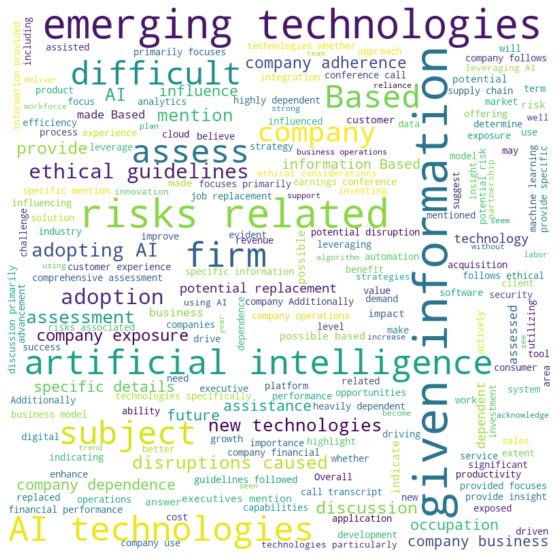}
\caption{Figure 5C. Word Cloud of \textit{AIRiskAssess}}
\end{figure}

\clearpage
\newpage

\onehalfspacing
\section*{Figure 6. Relative Importance of Risk Information}\phantomsection\label{fig6}
\begin{spacing}{0.9}
{\footnotesize\noindent This figure shows the relative importance of each risk measure over time. We set four-quarter rolling estimation windows. In 5A, we estimate $Implied\_Volatility_{it}  = \beta PRiskAssess_{it} + \gamma \textbf{X}_{it} + \delta_q + \delta_s + \varepsilon_{it}$, $Implied\_Volatility_{it}  = \beta CRiskAssess_{it} + \gamma \textbf{X}_{it} + \delta_q + \delta_s + \varepsilon_{it}$, and $Implied\_Volatility_{it}  = \beta AIRiskAssess_{it} + \gamma \textbf{X}_{it} + \delta_q + \delta_s + \varepsilon_{it}$ separately and plot the \textit{t}-values of each $\beta$. In 5B, we estimate $Implied\_Volatility_{it} =  \beta_1 PRiskAssess_{it} + \beta_2 CRiskAssess_{it} + \beta_3 AIRiskAssess_{it}  + \gamma \textbf{X}_{it} + \delta_q + \delta_s + \varepsilon_{it}$ and plot the \textit{t}-values of $\beta_1$, $\beta_2$, and $\beta_3$.} \textit{t}-values are clustered at the firm-level.
\end{spacing}

\begin{figure}[hbt!]
\centering
\captionsetup{labelformat=empty}
\includegraphics[width=130mm]{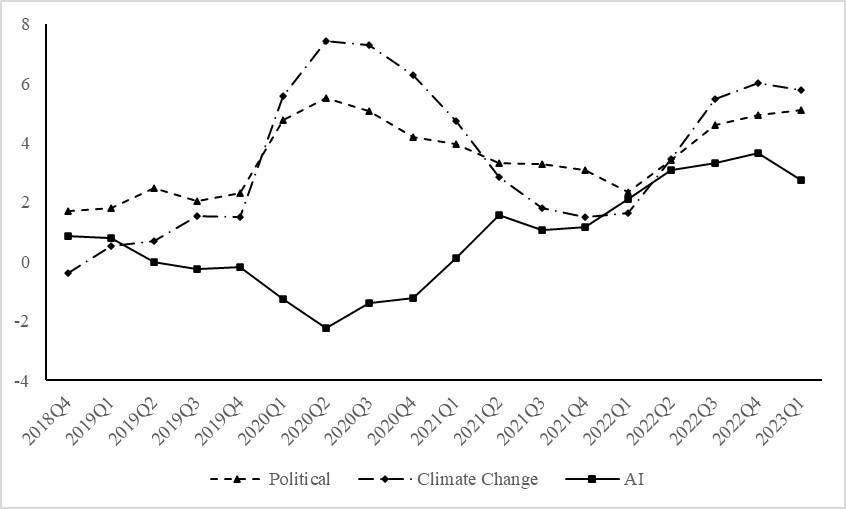}
\caption{Figure 6A. Separate Regressions}
\end{figure}

\begin{figure}[hbt!]
\centering
\captionsetup{labelformat=empty}
\includegraphics[width=130mm]{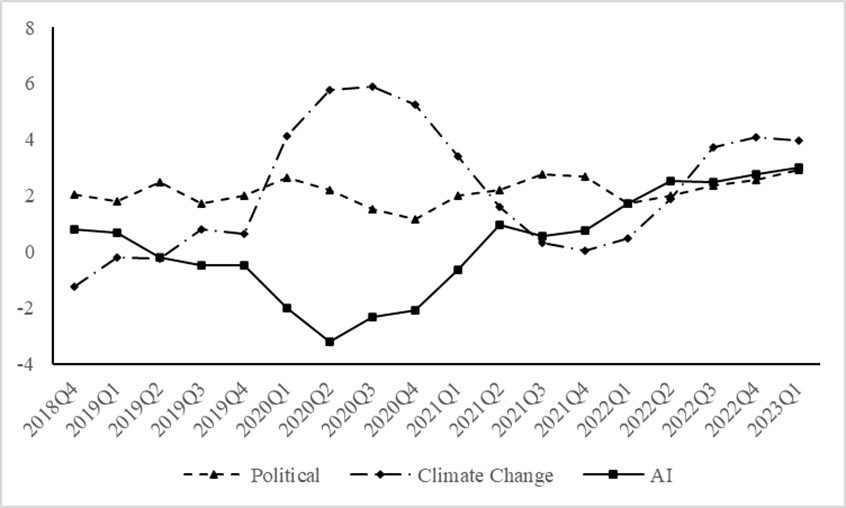}
\caption{Figure 6B. Regressions with All Three Risks}
\end{figure}

\clearpage
\newpage

\section*{Table 1. Descriptive Statistics}\phantomsection\label{t1}
\begin{spacing}{0.9}
{\footnotesize\noindent This table reports the descriptive statistics for the GPT-generated firm-level risk exposure variables and the key dependent variables for our full sample from 2018 to March 2023. Refer to Section 3 and Section 4 for variable descriptions. Panel A reports summary statistics. Panel B reports Pearson correlations among the variables. $PRB$ is an abbreviation for $PRiskBigram$, $PRS$ for $PRiskSum$, $PRA$ for $PRiskAssess$, $CRB$ for $CRiskBigram$, $CRS$ for $CRiskSum$, $CRA$ for $CRiskAssess$, $AIRS$ for $AIRiskSum$, and $AIRA$ for $AIRiskAssess$. }
\end{spacing}
\begin{table}[hbt!]
\centering
\begin{footnotesize}
\begin{tabular}{p{3cm}>{\centering}p{2.2cm}>{\centering}p{2.2cm}>{\centering}p{2.2cm}>{\centering}p{2.2cm}>{\centering\arraybackslash}p{2.2cm}} 
\hline
\multicolumn{6}{l}{\textbf{Panel A}. Summary Statistics}              \\ 
\hline
~                     & Mean  & Std   & P25   & P50   & P75    \\ 
\cline{2-6}
\textit{PRiskSum}     & 0.006 & 0.009 & 0.000 & 0.002 & 0.011  \\
\textit{PRiskAssess}  & 0.013 & 0.013 & 0.000 & 0.011 & 0.020  \\
\textit{CRiskSum}     & 0.002 & 0.005 & 0.000 & 0.000 & 0.002  \\
\textit{CRiskAssess}  & 0.006 & 0.009 & 0.000 & 0.001 & 0.009  \\
\textit{AIRiskSum}    & 0.001 & 0.004 & 0.000 & 0.000 & 0.000  \\
\textit{AIRiskAssess} & 0.004 & 0.006 & 0.000 & 0.000 & 0.007  \\
\textit{Implied Volatility }      & 0.488  & 0.268 & 0.300  & 0.415  & 0.591   \\
\textit{Abnormal Volatility } & -0.124 & 0.482 & -0.399 & -0.234 & -0.010  \\
\textit{Investments }             & 1.083  & 0.733 & 0.313  & 1.238  & 1.449   \\
$\textbf{1}\left(\text{\$ Lobby Amount}>0\right)$                       & 0.135  & 0.342 & 0.000  & 0.000  & 0.000   \\
$\textbf{1}\left(\text{\# Green Patent}>0\right)$                       & 0.121  & 0.326 & 0.000  & 0.000  & 0.000   \\
$\textbf{1}\left(\text{\# AI Patent}>0\right)$                       & 0.104  & 0.305 & 0.000  & 0.000  & 0.000 \\
\hline
\end{tabular}
\end{footnotesize}
\end{table}

\begin{table}[hbt!]
\centering
\begin{footnotesize}
\begin{tabular}{p{3.2cm}>{\centering}p{1.2cm}>{\centering}p{1.2cm}>{\centering}p{1.2cm}>{\centering}p{1.2cm}>{\centering}p{1.2cm}>{\centering}p{1.2cm}>{\centering}p{1.2cm}>{\centering\arraybackslash}p{1.2cm}} 
\hline
\multicolumn{9}{l}{\textbf{Panel B}. Correlations  \textit{~  ~  ~  ~  ~  ~  ~  ~}}                                                      \\ 
\hline
~             & \textit{PRB} & \textit{PRS} & \textit{PRA} & \textit{CRB} & \textit{CRS} & \textit{CRA} & \textit{AIRS} & \textit{AIRA}  \\ 
\cline{2-9}
\textit{PRB}  & 1.000        & ~            & ~            & ~            & ~            & ~            & ~             & ~              \\
\textit{PRS}  & 0.116        & 1.000        & ~            & ~            & ~            & ~            & ~             & ~              \\
\textit{PRA}  & 0.120        & 0.742        & 1.000        & ~            & ~            & ~            & ~             & ~              \\
\textit{CRB}  & 0.011        & 0.139        & 0.150        & 1.000        & ~            & ~            & ~             & ~              \\
\textit{CRS}  & 0.036        & 0.349        & 0.364        & 0.246        & 1.000        & ~            & ~             & ~              \\
\textit{CRA}  & 0.047        & 0.415        & 0.483        & 0.260        & 0.661        & 1.000        & ~             & ~              \\
\textit{AIRS} & 0.004        & 0.050        & 0.060        & -0.014       & 0.071        & 0.049        & 1.000         & ~              \\
\textit{AIRA} & -0.013       & 0.044        & 0.088        & -0.006       & 0.069        & 0.055        & 0.512         & 1.000          \\
\hline
\end{tabular}
\end{footnotesize}
\end{table}

\clearpage
\newpage

\section*{Table 2. Variance Decomposition}\phantomsection\label{t2}
\begin{spacing}{0.9}
{\footnotesize\noindent This table the variance decomposition results for each risk exposure variable for our full sample from 2018 to March 2023. We report political risk exposure measures in Panel A, climate change risk exposure measures in Panel B, and AI-related risk exposure measures in Panel C. The first subpanel reports incremental R-squared values by adding time, industry, and time$\times$industry fixed effects. The second subpanel reports R-squared values by adding firm fixed effects.}
\end{spacing}

\begin{table}[hbt!]
\centering
\begin{footnotesize}
\begin{tabular}{p{5cm}>{\centering}p{3.3cm}>{\centering}p{3.3cm}>{\centering\arraybackslash}p{3.3cm}} 
\hline
\multicolumn{4}{l}{\textbf{Panel A}. Political Risk ~ ~ ~}                                              \\ 
\hline
~                              & \textit{ PRiskSum } & \textit{ PRiskAssess } & \textit{ PRiskBigram }  \\ 
\hline
\textit{Panel A1}.                      & Incremental
  R\textsuperscript{2}    & Incremental
  R\textsuperscript{2}       & Incremental
  R\textsuperscript{2}        \\ 
\cline{2-4}
Time FE                        & 1.94\%              & 4.51\%                 & 3.20\%                  \\
Industry FE                    & 3.51\%              & 4.64\%                 & 3.50\%                  \\
Time$\times$Industry
  FE             & 2.79\%              & 2.82\%                 & 1.96\%                  \\
Implied Firm
  Level Variation & 91.76\%             & 88.03\%                & 91.34\%                 \\
Sum                            & 100.00\%            & 100.00\%               & 100.00\%                \\ 
\hline
~                              & ~                   & ~                      & ~                       \\ 
\hline
\textit{Panel A2}.                      & R\textsuperscript{2}                  & R\textsuperscript{2}                     & R\textsuperscript{2}                      \\ 
\cline{2-4}
Firm FE                        & 32.09\%             & 35.58\%                & 30.79\%                 \\
Remaining
  Variation          & 67.91\%             & 64.42\%                & 69.21\%                 \\
Sum                            & 100.00\%            & 100.00\%               & 100.00\%                \\
\hline
\end{tabular}
\end{footnotesize}
\end{table}

\begin{table}[hbt!]
\centering
\begin{footnotesize}
\begin{tabular}{p{5cm}>{\centering}p{3.3cm}>{\centering}p{3.3cm}>{\centering\arraybackslash}p{3.3cm}} 
\hline
\multicolumn{4}{l}{\textbf{Panel B}. Climate Change Risk ~}                                             \\ 
\hline
~                              & \textit{ CRiskSum } & \textit{ CRiskAssess } & \textit{ CRiskBigram }  \\ 
\hline
\textit{Panel B1}.                      & Incremental
  R\textsuperscript{2}    & Incremental
  R\textsuperscript{2}       & Incremental
  R\textsuperscript{2}        \\ 
\cline{2-4}
Time FE                        & 4.05\%              & 5.82\%                 & 0.30\%                  \\
Industry FE                    & 5.82\%              & 8.44\%                 & 7.86\%                  \\
Time$\times$Industry
  FE             & 4.10\%              & 3.48\%                 & 2.36\%                  \\
Implied Firm
  Level Variation & 86.03\%             & 82.26\%                & 89.48\%                 \\
Sum                            & 100.00\%            & 100.00\%               & 100.00\%                \\ 
\hline
~                              & ~                   & ~                      & ~                       \\ 
\hline
\textit{Panel B2}.                      & R\textsuperscript{2}                  & R\textsuperscript{2}                     & R\textsuperscript{2}                      \\ 
\cline{2-4}
Firm FE                        & 24.13\%             & 32.28\%                & 32.36\%                 \\
Remaining
  Variation          & 75.87\%             & 67.72\%                & 67.64\%                 \\
Sum                            & 100.00\%            & 100.00\%               & 100.00\%                \\
\hline
\end{tabular}
\end{footnotesize}
\end{table}

\begin{table}[hbt!]
\centering
\begin{footnotesize}
\begin{tabular}{p{5cm}>{\centering}p{3.3cm}>{\centering}p{3.3cm}>{\centering\arraybackslash}p{3.3cm}} 
\hline
\multicolumn{4}{l}{\textbf{Panel C}. AI Exposure Risk ~ ~}                       \\ 
\hline
~                              & \textit{ AIRiskSum } & \textit{ AIRiskAssess }  & \textit{ AIRiskBigram } \\ 
\hline
\textit{Panel C1}.            & Incremental
  R\textsuperscript{2}     & Incremental
  R\textsuperscript{2}      & Incremental
  R\textsuperscript{2}  \\ 
\cline{2-4}
Time FE                        & 1.07\%               & 5.00\%       & -           \\
Industry FE                    & 3.37\%               & 9.67\%      & -             \\
Time$\times$Industry
  FE             & 2.99\%               & 2.73\%                 & -  \\
Implied Firm
  Level Variation & 92.57\%              & 82.60\%              & -    \\
Sum                            & 100.00\%             & 100.00\%          & ~       \\ 
\hline
~                              & ~                    & ~                    & ~    \\ 
\hline
\textit{Panel C2. }           & R\textsuperscript{2}                   & R\textsuperscript{2}  & R\textsuperscript{2}                     \\ 
\cline{2-4}
Firm FE                        & 21.31\%              & 31.31\%       &   -         \\
Remaining
  Variation          & 78.69\%              & 68.69\%                & -  \\
Sum                            & 100.00\%             & 100.00\%     & -            \\
\hline
\end{tabular}
\end{footnotesize}
\end{table}
\clearpage
\newpage

\section*{Table 3. Political Risk Exposure and Volatility}\phantomsection\label{t3}
\begin{spacing}{0.9}
{\footnotesize\noindent This table reports the association between firm-level political risk exposure measures and volatility using our main sample period from 2018 to 2021. We use industry and time fixed effects in Panel A, and industry$\times$time fixed effects in Panel B. All continuous variables are winsorized at 1\% and 99\% levels. Standard errors are clustered at firm-level. ***, **, and * denote significance at 1\%, 5\%, and 10\% levels, respectively.}
\end{spacing}

\begin{table}[hbt!]
\centering
\begin{scriptsize}
\begin{tabular}{p{3cm}>{\centering}p{2.2cm}>{\centering}p{2.2cm}>{\centering}p{2.2cm}>{\centering}p{2.2cm}>{\centering\arraybackslash}p{2.2cm}} 
\hline
\multicolumn{6}{l}{\textbf{Panel A}. Industry and Time Fixed Effects}   \\ 
\hline
\multicolumn{6}{l}{Dep Var = Implied Volatility}                                        \\ 
\cline{2-6}
~                      & (1)      & (2)      & (3)      & (4)      & (5)        \\ 
\hline
\textit{ PRiskSum }    & 0.888*** & ~        & 0.857*** & ~        & 0.160      \\
\textit{ ~ }           & (4.50)   & ~        & (4.37)   & ~        & (0.73)     \\
\textit{ PRiskAssess } & ~        & 0.777*** & ~        & 0.757*** & 0.676***   \\
\textit{ ~ }           & ~        & (5.08)   & ~        & (4.97)   & (3.72)     \\
\textit{ PRiskBigram } & ~        & ~        & 0.015    & 0.014    & 0.014      \\
\textit{ ~ }           & ~        & ~        & (1.54)   & (1.43)   & (1.40)     \\
Controls               & Yes      & Yes      & Yes      & Yes      & Yes        \\
Fixed Effects & Time \& Ind & Time \& Ind & Time \& Ind &Time \& Ind &Time \& Ind \\
N                      & 35003    & 35003    & 35003    & 35003    & 35003      \\
Adjusted R\textsuperscript{2}  & 0.477    & 0.477    & 0.477    & 0.477    & 0.477      \\ 
\hline

\multicolumn{6}{l}{Dep Var =
  Abnormal Volatility}                                           \\ 
\cline{2-6}
~                      & (1)      & (2)      & (3)      & (4)      & (5)        \\ 
\hline
\textit{ PRiskSum }    & 1.576*** & ~        & 1.574*** & ~        & -1.031***  \\
\textit{ ~ }           & (6.55)   & ~        & (6.54)   & ~        & (-2.96)    \\
\textit{ PRiskAssess } & ~        & 1.987*** & ~        & 1.995*** & 2.523***   \\
\textit{ ~ }           & ~        & (11.42)  & ~        & (11.44)  & (10.00)    \\
\textit{ PRiskBigram } & ~        & ~        & 0.001    & -0.006   & -0.004     \\
\textit{ ~ }           & ~        & ~        & (0.07)   & (-0.48)  & (-0.34)    \\
Controls               & Yes      & Yes      & Yes      & Yes      & Yes        \\
Fixed Effects & Time \& Ind & Time \& Ind & Time \& Ind &Time \& Ind &Time \& Ind \\
N                      & 39276    & 39276    & 39276    & 39276    & 39276      \\
Adjusted R\textsuperscript{2}                 & 0.353    & 0.355    & 0.353    & 0.355    & 0.355      \\
\hline
\end{tabular}
\end{scriptsize}
\end{table}

\begin{table}[hbt!]
\centering
\begin{scriptsize}
\begin{tabular}{p{3cm}>{\centering}p{2.2cm}>{\centering}p{2.2cm}>{\centering}p{2.2cm}>{\centering}p{2.2cm}>{\centering\arraybackslash}p{2.2cm}} 
\hline
\multicolumn{6}{l}{\textbf{Panel B}. Industry$\times$Time Fixed Effects}   \\ 
\hline
\multicolumn{6}{l}{Dep Var = Implied Volatility}                                        \\ 
\cline{2-6}
~                      & (1)      & (2)      & (3)      & (4)      & (5)       \\ 
\hline
\textit{ PRiskSum }    & 0.737*** & ~        & 0.704*** & ~        & 0.154     \\
\textit{ ~ }           & (3.64)   & ~        & (3.49)   & ~        & (0.69)    \\
\textit{ PRiskAssess } & ~        & 0.635*** & ~        & 0.614*** & 0.535***  \\
\textit{ ~ }           & ~        & (4.07)   & ~        & (3.95)   & (2.92)    \\
\textit{ PRiskBigram } & ~        & ~        & 0.016    & 0.015    & 0.015     \\
\textit{ ~ }           & ~        & ~        & (1.60)   & (1.52)   & (1.50)    \\
Controls               & Yes     & Yes      & Yes      & Yes      & Yes      \\
Fixed Effects & Time $\times$ Ind & Time $\times$ Ind & Time $\times$ Ind &Time $\times$ Ind &Time $\times$ Ind \\
N                      & 35003    & 35003    & 35003    & 35003    & 35003      \\
Adjusted R\textsuperscript{2}  & 0.477    & 0.477    & 0.477    & 0.477    & 0.477      \\ 
\hline

\multicolumn{6}{l}{Dep Var =
  Abnormal Volatility}                                           \\ 
\cline{2-6}
~                      & (1)      & (2)      & (3)      & (4)      & (5)       \\ 
\hline
\textit{ PRiskSum }    & 1.559*** & ~        & 1.559*** & ~        & -0.659**  \\
\textit{ ~ }           & (6.62)   & ~        & (6.61)   & ~        & (-1.96)   \\
\textit{ PRiskAssess } & ~        & 1.808*** & ~        & 1.815*** & 2.152***  \\
\textit{ ~ }           & ~        & (10.51)  & ~        & (10.52)  & (8.75)    \\
\textit{ PRiskBigram } & ~        & ~        & 0.000    & -0.005   & -0.004    \\
\textit{ ~ }           & ~        & ~        & (0.02)   & (-0.43)  & (-0.34)   \\
Controls               & Yes     & Yes      & Yes      & Yes      & Yes      \\
Fixed Effects & Time $\times$ Ind & Time $\times$ Ind & Time $\times$ Ind &Time $\times$ Ind &Time $\times$ Ind \\
N                      & 35003    & 35003    & 35003    & 35003    & 35003      \\
Adjusted R\textsuperscript{2}  & 0.477    & 0.477    & 0.477    & 0.477    & 0.477      \\ 
\hline
\end{tabular}
\end{scriptsize}
\end{table}

\clearpage
\newpage

\section*{Table 4. Climate Change Risk Exposure and Volatility}\phantomsection\label{t4}
\begin{spacing}{0.9}
{\footnotesize\noindent This table reports the association between firm-level climate change risk exposure measures and volatility using our main sample period from 2018 to 2021. We use industry and time fixed effects in Panel A, and industry$\times$time fixed effects in Panel B. All continuous variables are winsorized at 1\% and 99\% levels. Standard errors are clustered at firm-level. ***, **, and * denote significance at 1\%, 5\%, and 10\% levels, respectively.}
\end{spacing}

\begin{table}[hbt!]
\centering
\begin{scriptsize}
\begin{tabular}{p{3cm}>{\centering}p{2.2cm}>{\centering}p{2.2cm}>{\centering}p{2.2cm}>{\centering}p{2.2cm}>{\centering\arraybackslash}p{2.2cm}} 
\hline
\multicolumn{6}{l}{\textbf{Panel A}. Industry and Time Fixed Effects}   \\ 
\hline
\multicolumn{6}{l}{Dep Var = Implied Volatility}                                        \\ 
\cline{2-6}
~                     & (1)     & (2)      & (3)     & (4)      & (5)       \\ 
\hline
\textit{ CRiskSum }    & 0.807** & ~        & 0.769** & ~        & -0.576*   \\
\textit{ ~ }          & (2.51)  & ~        & (2.42)  & ~        & (-1.72)   \\
\textit{ CRiskAssess } & ~       & 1.075*** & ~       & 1.074*** & 1.281***  \\
\textit{ ~ }          & ~       & (4.51)   & ~       & (4.53)   & (4.74)    \\
\textit{ CRiskBigram } & ~       & ~        & 0.069   & -0.316   & -0.251    \\
\textit{ ~ }          & ~       & ~        & (0.07)  & (-0.34)  & (-0.27)   \\
Controls               & Yes      & Yes      & Yes      & Yes      & Yes        \\
Fixed Effects & Time \& Ind & Time \& Ind & Time \& Ind &Time \& Ind &Time \& Ind \\
N                     & 35003   & 35003    & 35003   & 35003    & 35003     \\
Adjusted R\textsuperscript{2}                & 0.476   & 0.477    & 0.476   & 0.477    & 0.477     \\
\hline

\multicolumn{6}{l}{Dep Var =
  Abnormal Volatility}                                           \\ 
\cline{2-6}
~                      & (1)      & (2)      & (3)      & (4)      & (5)        \\ 
\hline
\textit{ CRiskSum }    & 1.693*** & ~        & 1.502*** & ~        & -1.401**  \\
\textit{ ~ }           & (3.72)   & ~        & (3.28)   & ~        & (-2.41)   \\
\textit{ CRiskAssess } & ~        & 2.332*** & ~        & 2.269*** & 2.752***  \\
\textit{ ~ }           & ~        & (9.32)   & ~        & (8.85)   & (8.49)    \\
\textit{ CRiskBigram } & ~        & ~        & 1.678**  & 0.807    & 0.967     \\
\textit{ ~ }           & ~        & ~        & (2.27)   & (1.08)   & (1.29)    \\
Controls               & Yes      & Yes      & Yes      & Yes      & Yes        \\
Fixed Effects & Time \& Ind & Time \& Ind & Time \& Ind &Time \& Ind &Time \& Ind \\
N                      & 39276    & 39276    & 39276    & 39276    & 39276      \\
Adjusted R\textsuperscript{2}                 & 0.353    & 0.354    & 0.353    & 0.354    & 0.354      \\
\hline
\end{tabular}
\end{scriptsize}
\end{table}

\begin{table}[hbt!]
\centering
\begin{scriptsize}
\begin{tabular}{p{3cm}>{\centering}p{2.2cm}>{\centering}p{2.2cm}>{\centering}p{2.2cm}>{\centering}p{2.2cm}>{\centering\arraybackslash}p{2.2cm}} 
\hline
\multicolumn{6}{l}{\textbf{Panel B}. Industry$\times$Time Fixed Effects}   \\ 
\hline
\multicolumn{6}{l}{Dep Var = Implied Volatility}                                        \\ 
\cline{2-6}
~                     & (1)     & (2)      & (3)     & (4)      & (5)       \\ 
\hline
\textit{ CRiskSum }    & 0.676** & ~        & 0.660** & ~        & -0.500    \\
\textit{ ~ }          & (2.00)  & ~        & (1.99)  & ~        & (-1.45)   \\
\textit{ CRiskAssess } & ~       & 0.915*** & ~       & 0.924*** & 1.102***  \\
\textit{ ~ }          & ~       & (3.70)   & ~       & (3.76)   & (3.97)    \\
\textit{ CRiskBigram } & ~       & ~        & -0.077  & -0.396   & -0.344    \\
\textit{ ~ }          & ~       & ~        & (-0.08) & (-0.41)  & (-0.35)   \\
Controls               & Yes     & Yes      & Yes      & Yes      & Yes      \\
Fixed Effects & Time $\times$ Ind & Time $\times$ Ind & Time $\times$ Ind &Time $\times$ Ind &Time $\times$ Ind \\
N                      & 35003    & 35003    & 35003    & 35003    & 35003      \\
Adjusted R\textsuperscript{2}                & 0.489   & 0.489    & 0.489   & 0.489    & 0.489     \\
\hline

\multicolumn{6}{l}{Dep Var =
  Abnormal Volatility}                                           \\ 
\cline{2-6}
~                     & (1)      & (2)      & (3)      & (4)      & (5)        \\ 
\hline
\textit{ CRiskSum }    & 1.262*** & ~        & 1.019**  & ~        & -1.906***  \\
\textit{ ~ }          & (2.91)   & ~        & (2.33)   & ~        & (-3.45)    \\
\textit{ CRiskAssess } & ~        & 2.216*** & ~        & 2.116*** & 2.768***   \\
\textit{ ~ }          & ~        & (9.20)   & ~        & (8.58)   & (8.90)     \\
\textit{ CRiskBigram } & ~        & ~        & 2.253*** & 1.355*   & 1.557**    \\
\textit{ ~ }          & ~        & ~        & (3.11)   & (1.87)   & (2.15)     \\
Controls               & Yes     & Yes      & Yes      & Yes      & Yes      \\
Fixed Effects & Time $\times$ Ind & Time $\times$ Ind & Time $\times$ Ind &Time $\times$ Ind &Time $\times$ Ind \\
N                      & 35003    & 35003    & 35003    & 35003    & 35003      \\
Adjusted R\textsuperscript{2}                & 0.455    & 0.457    & 0.456    & 0.457    & 0.457      \\
\hline
\end{tabular}
\end{scriptsize}
\end{table}

\clearpage
\newpage

\section*{Table 5. AI-related Risk Exposure and Volatility}\phantomsection\label{t5}
\begin{spacing}{0.9}
{\footnotesize\noindent This table reports the association between firm-level AI-related risk exposure measures and volatility using our main sample period from 2018 to 2021. We use industry and time fixed effects in Panel A, and industry$\times$time fixed effects in Panel B. All continuous variables are winsorized at 1\% and 99\% levels. Standard errors are clustered at firm-level. ***, **, and * denote significance at 1\%, 5\%, and 10\% levels, respectively.}
\end{spacing}

\begin{table}[hbt!]
\centering
\begin{footnotesize}
\begin{tabular}{p{2.2cm}>{\centering}p{1.9cm}>{\centering}p{1.9cm}>{\centering}p{1.9cm}>{\centering}p{1.9cm}>{\centering}p{1.9cm}>{\centering\arraybackslash}p{1.9cm}} 
\hline
\multicolumn{7}{l}{\textbf{Panel A}. Industry and Time Fixed Effects~ ~ ~ ~ ~ ~ ~~}                                       \\ 
\hline
~                     & \multicolumn{3}{c}{Implied
  Volatility ~ ~} & \multicolumn{3}{c}{Abnormal~Volatility ~ ~}  \\ 
\cline{2-7}
~                     & (1)     & (2)     & (3)                      & (4)      & (5)    & (6)                            \\ 
\hline
\textit{ AIRiskSum}    & -0.457  & ~       & -0.444                   & 1.738*** & ~      & 1.853***                       \\
~                     & (-1.38) & ~       & (-1.34)                  & (3.19)   & ~      & (3.01)                         \\
\textit{ AIRiskAssess} & ~       & -0.170  & -0.019                   & ~        & 0.459  & -0.166                         \\
~                     & ~       & (-0.60) & (-0.07)                  & ~        & (1.22) & (-0.39)                        \\
Controls              & Yes     & Yes     & Yes                      & Yes      & Yes    & Yes                            \\
Time FE               & Yes     & Yes     & Yes                      & Yes      & Yes    & Yes                            \\
Industry FE           & Yes     & Yes     & Yes                      & Yes      & Yes    & Yes                            \\
Time*Ind FE           & No      & No      & No                       & No       & No     & No                             \\
N                     & 35003   & 35003   & 35003                    & 39276    & 39276  & 39276                          \\
Adjusted R\textsuperscript{2}                & 0.476   & 0.476   & 0.476                    & 0.353    & 0.352  & 0.353                          \\

\hline
\multicolumn{7}{l}{\textbf{Panel B}. Industry$\times$Time Fixed Effects~ ~ ~ ~ ~ ~ ~~}                                          \\ 
\hline
~                     & \multicolumn{3}{c}{Implied
  Volatility ~ ~}    & \multicolumn{3}{c}{Abnormal~Volatility ~ ~}  \\ 
\cline{2-7}
~                     & (1)                         & (2)     & (3)     & (4)     & (5)    & (6)                             \\ 
\hline
\textit{ AIRiskSum}    & -0.520 & ~       & -0.477  & 1.153** & ~      & 1.853***                        \\
~                     & (-1.56)                     & ~       & (-1.42) & (2.25)  & ~      & (2.36)                          \\
\textit{ AIRiskAssess} & ~                           & -0.224  & -0.061  & ~       & 0.166  & -0.292                          \\
~                     & ~                           & (-0.79) & (-0.20) & ~       & (0.46) & (-0.72)                         \\
Controls              & Yes                         & Yes     & Yes     & Yes     & Yes    & Yes                             \\
Time FE               & No                          & No      & No      & No      & No     & No                              \\
Industry FE           & No                          & No      & No      & No      & No     & No                              \\
Time*Ind FE           & Yes                         & Yes     & Yes     & Yes     & Yes    & Yes                             \\
N                     & 35003                       & 35003   & 35003   & 39276   & 39276  & 39276                           \\
Adjusted R\textsuperscript{2}                & 0.489                       & 0.489   & 0.489   & 0.455   & 0.455  & 0.455                           \\
\hline
\end{tabular}
\end{footnotesize}
\end{table}



\clearpage
\newpage

\section*{Table 6. Out-of-Sample Analysis: Capital Market Consequences}\phantomsection\label{t6}
\begin{spacing}{0.9}
{\footnotesize\noindent This table repeats the analysis of Tables 3, 4, and 5 with a sample from 2022 to March 2023. We report political risk exposure measures in Panel A, climate change risk exposure measures in Panel B, and AI-related risk exposure measures in Panel C. We use industry and time fixed effects. All continuous variables are winsorized at 1\% and 99\% levels. Standard errors are clustered at firm-level. ***, **, and * denote significance at 1\%, 5\%, and 10\% levels, respectively.}
\end{spacing}

\begin{table}[hbt!]
\centering
\begin{footnotesize}
\begin{tabular}{p{4.4cm}>{\centering}p{2.5cm}>{\centering}p{2.5cm}>{\centering}p{2.5cm}>{\centering\arraybackslash}p{2.5cm}} 
\hline
\multicolumn{5}{l}{\textbf{Panel A}. Political Risks ~ ~ ~ ~}                                                  \\ 
\hline
~                      & \multicolumn{2}{c}{Implied
  Volatility ~} & \multicolumn{2}{c}{Abnormal
  Volatility ~}  \\ 
\cline{2-5}
~                      & (1)      & (2)                             & (3)    & (4)                             \\ 
\hline
\textit{ PRiskSum }    & 1.716*** & ~                               & 0.430  & ~                               \\
~                      & (5.04)   & ~                               & (1.09) & ~                               \\
\textit{ PRiskAssess } & ~        & 1.246***                        & ~      & 0.941***                        \\
~                      & ~        & (5.12)                          & ~      & (3.44)                          \\
Controls               & Yes      & Yes                             & Yes    & Yes                             \\
Time FE                & Yes      & Yes                             & Yes    & Yes                             \\
Industry FE            & Yes      & Yes                             & Yes    & Yes                             \\
N                      & 9923     & 9923                            & 9246   & 9246                            \\
Adjusted R\textsuperscript{2}                 & 0.423    & 0.423                           & 0.115  & 0.117                           \\
\hline
\multicolumn{5}{l}{\textbf{Panel B}. Climate Change Risks ~ ~ ~ ~}                                             \\ 
\hline
~                      & \multicolumn{2}{c}{Implied
  Volatility ~} & \multicolumn{2}{c}{Abnormal Volatility ~}  \\ 
\cline{2-5}
~                      & (1)      & (2)                             & (3)    & (4)                             \\ 
\hline
\textit{ CRiskSum }    & 2.072*** & ~                               & 0.276  & ~                               \\
~                      & (4.26)   & ~                               & (0.46) & ~                               \\
\textit{ CRiskAssess } & ~        & 2.135***                        & ~      & 1.494***                        \\
~                      & ~        & (6.14)                          & ~      & (3.76)                          \\
Controls               & Yes      & Yes                             & Yes    & Yes                             \\
Time FE                & Yes      & Yes                             & Yes    & Yes                             \\
Industry FE            & Yes      & Yes                             & Yes    & Yes                             \\
N                      & 9923     & 9923                            & 9246   & 9246                            \\
Adjusted R\textsuperscript{2}                 & 0.421    & 0.424                           & 0.115  & 0.117                           \\
\hline
\multicolumn{5}{l}{\textbf{Panel C}. AI-Related Risk~ ~ ~ ~~}                                                    \\ 
\hline
~                       & \multicolumn{2}{c}{Implied
  Volatility ~} & \multicolumn{2}{c}{Abnormal
  Volatility ~}  \\ 
\cline{2-5}
~                       & (1)    & (2)                               & (3)      & (4)                           \\ 
\hline
\textit{ AIRiskSum }    & 0.775* & ~                                 & 2.281*** & ~                             \\
~                       & (1.85) & ~                                 & (3.69)   & ~                             \\
\textit{ AIRiskAssess } & ~      & 1.166***                          & ~        & 1.259***                      \\
~                       & ~      & (3.53)                            & ~        & (2.76)                        \\
Controls                & Yes    & Yes                               & Yes      & Yes                           \\
Time FE                 & Yes    & Yes                               & Yes      & Yes                           \\
Industry FE             & Yes    & Yes                               & Yes      & Yes                           \\
N                       & 9923   & 9923                              & 9246     & 9246                          \\
Adjusted R\textsuperscript{2}                  & 0.420  & 0.421                             & 0.116    & 0.116                         \\
\hline
\end{tabular}
\end{footnotesize}
\end{table}

\clearpage
\newpage

\section*{Table 7. Risk Exposure and Investments}\phantomsection\label{t7}
\begin{spacing}{0.9}
{\footnotesize\noindent This table reports the association between firm-level risk exposure variables and capital investments using our main sample period from 2018 to 2021. We report political risk exposure measures in Panel A, climate change risk exposure measures in Panel B, and AI-related risk exposure measures in Panel C. We use capital expenditure scaled by recursive total capital as a dependent variable. We use industry and time fixed effects. All continuous variables are winsorized at 1\% and 99\% levels. Standard errors are clustered at firm-level. ***, **, and * denote significance at 1\%, 5\%, and 10\% levels, respectively.}
\end{spacing}

\begin{table}[hbt!]
\centering
\begin{footnotesize}
\begin{tabular}{p{3cm}>{\centering}p{2.2cm}>{\centering}p{2.2cm}>{\centering}p{2.2cm}>{\centering}p{2.2cm}>{\centering\arraybackslash}p{2.2cm}} 
\hline
\multicolumn{6}{l}{\textbf{Panel A}.
  Political Risk ~}                                             \\ 
\hline
Dep Var                & \multicolumn{5}{c}{Investment}                          \\ 
\cline{2-6}
~                     & (1)      & (2)       & (3)      & (4)       & (5)        \\ 
\hline
\textit{ PRiskSum }    & -0.792** & ~         & -0.792** & ~         & 0.317      \\
\textit{ ~ }          & (-2.08)  & ~         & (-2.06)  & ~         & (0.58)     \\
\textit{ PRiskAssess } & ~        & -0.918*** & ~        & -0.922*** & -1.081***  \\
\textit{ ~ }          & ~        & (-3.33)   & ~        & (-3.32)   & (-2.72)    \\
\textit{ PRiskBigram } & ~        & ~         & -0.000   & 0.003     & 0.002      \\
\textit{ ~ }          & ~        & ~         & (-0.01)  & (0.13)    & (0.11)     \\
Controls                   & Yes      & Yes       & Yes      & Yes       & Yes        \\
Time FE                     & Yes      & Yes       & Yes      & Yes       & Yes        \\
Ind FE                     & Yes      & Yes       & Yes      & Yes       & Yes        \\
N                        & 35615    & 35615     & 35615    & 35615     & 35615      \\
Adjusted R\textsuperscript{2}                   & 0.279    & 0.280     & 0.279    & 0.280     & 0.280      \\
\hline

\multicolumn{6}{l}{\textbf{Panel B}. Climate Change Risk}                                 \\ 
\hline
Dep Var                & \multicolumn{5}{c}{Investment}                          \\ 
\cline{2-6}
~                     & (1)     & (2)       & (3)     & (4)       & (5)        \\ 
\hline
\textit{ CRiskSum }    & -0.606  &    & -0.524  & ~         & 0.924      \\
\textit{ ~ }          & (-0.90) &    & (-0.77) & ~         & (1.08)     \\
\textit{ CRiskAssess } & ~       & -1.077*** & ~       & -1.059** & -1.375***  \\
\textit{ ~ }          & ~       & (-2.62)   & ~       & (-2.54)   & (-2.64)    \\
\textit{ CRiskBigram } & ~       & ~         & -0.865  & -0.392    & -0.502      \\
\textit{ ~ }          & ~       & ~         & (-0.86) & (-0.39)    & (-0.50)     \\ 
Controls                    & Yes     & Yes       & Yes     & Yes       & Yes        \\
Time FE                  & Yes     & Yes       & Yes     & Yes       & Yes        \\
Ind FE                    & Yes     & Yes       & Yes     & Yes       & Yes        \\
N                        & 35615   & 35615     & 35615   & 35615     & 35615      \\
Adjusted R\textsuperscript{2}                  & 0.279   & 0.280     & 0.279   & 0.279     & 0.279      \\
\hline

\multicolumn{6}{l}{\textbf{Panel C}. AI-Related Risk}                                 \\
\hline
Dep Var                & \multicolumn{5}{c}{Investment}                          \\ 
\cline{2-6}
~                      & (1)      & (2)       & (3)      & (4)       & (5)        \\ 
\hline

\textit{ AIRiskSum}    & 0.694  & ~      & -   & ~   & 0.231   \\
~             & (0.84) & ~      & ~   & ~   & (0.25)  \\
\textit{ AIRiskAssess}   & ~      & 0.751  & ~   & -   & 0.674   \\
~               & ~      & (1.28) & ~   & ~   & (1.02)  \\
\textit{ AIRiskBigram}    & ~      & ~      & -   & -   & -       \\
~               & ~      & ~      & ~   & ~   & ~       \\
Controls        & Yes    & Yes    & ~   & ~   & Yes     \\
Time FE         & Yes    & Yes    & ~   & ~   & Yes     \\
Ind FE         & Yes    & Yes    & ~   & ~   & Yes     \\
N               & 35615  & 35615  & ~   & ~   & 35615   \\
Adjusted R\textsuperscript{2}          & 0.279  & 0.280  & ~   & ~   & 0.279   \\
\hline
\end{tabular}
\end{footnotesize}
\end{table}

\clearpage
\newpage

\section*{Table 8. Risk Exposure and Firm Responses}\phantomsection\label{t8}
\begin{spacing}{0.9}
{\footnotesize\noindent This table reports how firms respond to different risk exposures using our main sample period from 2018 to 2021. We report political risk exposure measures in Panel A, climate change risk exposure measures in Panel B, and AI-related risk exposure measures in Panel C. For Panel A, we use lobbying activity indicator. For Panel B, we use green patent filing indicator and, for Panel C, we use AI patent filing indicator. We use industry and time fixed effects. All continuous variables are winsorized at 1\% and 99\% levels. Standard errors are clustered at firm-level. ***, **, and * denote significance at 1\%, 5\%, and 10\% levels, respectively.}
\end{spacing}

\begin{table}[hbt!]
\centering
\begin{footnotesize}
\begin{tabular}{p{3cm}>{\centering}p{2.2cm}>{\centering}p{2.2cm}>{\centering}p{2.2cm}>{\centering}p{2.2cm}>{\centering\arraybackslash}p{2.2cm}} 
\hline
\multicolumn{6}{l}{\textbf{Panel A}.
  Political Risk ~}                                             \\ 
\hline
Dep Var                & \multicolumn{5}{c}{$\textbf{1}\left(\text{\$ Lobby Amount}>0\right)$}                          \\ 
\cline{2-6}
~                     & (1)      & (2)       & (3)      & (4)       & (5)        \\ 
\hline
\textit{PRiskSum }    & 0.057  & ~       & 0.142   & ~      & -1.258***  \\
\textit{ ~ }          & (0.18) & ~       & (0.46)  & ~      & (-4.19)    \\
\textit{PRiskAssess } & ~      & 0.489** & ~       & 0.437* & 1.079***  \\
\textit{ ~ }          & ~      & (2.04)  & ~       & (1.84) & (4.21)    \\
\textit{PRiskBigram } & ~      & ~       & 0.041**  & 0.037*  & 0.039*     \\
\textit{ ~ }          & ~      & ~       & (2.04) & (1.83) & (1.94)    \\
Controls                   & Yes      & Yes       & Yes      & Yes       & Yes        \\
Time FE                     & Yes      & Yes       & Yes      & Yes       & Yes        \\
Ind FE                     & Yes      & Yes       & Yes      & Yes       & Yes        \\
N                        & 39937   & 39937   & 39937   & 39937  & 39937      \\
Adjusted R\textsuperscript{2}                & 0.191   & 0.191   & 0.191   & 0.191  & 0.192   \\
\hline

\multicolumn{6}{l}{\textbf{Panel B}. Climate Change Risk}                                 \\ 
\hline
Dep Var                & \multicolumn{5}{c}{$\textbf{1}\left(\text{\# Green Patent}>0\right)$}                          \\ 
\cline{2-6}
~                     & (1)    & (2)     & (3)    & (4)     & (5)       \\ 
\hline
\textit{ CRiskSum }    & 0.271  & ~       & 0.089  & ~       & -0.757*   \\
\textit{ ~ }          & (0.64) & ~       & (0.22) & ~       & (-1.77)   \\
\textit{ CRiskAssess } & ~      & 0.636** & ~      & 0.542** & 0.802***  \\
\textit{ ~ }          & ~      & (2.25)  & ~      & (2.01)  & (2.70)    \\
\textit{ CRiskBigram } & ~      & ~       & 2.141  & 1.847   & 1.934     \\
\textit{ ~ }          & ~      & ~       & (1.47) & (1.28)  & (1.34)    \\
Controls                    & Yes     & Yes       & Yes     & Yes       & Yes        \\
Time FE                  & Yes     & Yes       & Yes     & Yes       & Yes        \\
Ind FE                    & Yes     & Yes       & Yes     & Yes       & Yes        \\
N                        & 39937   & 39937   & 39937   & 39937  & 39937      \\
Adjusted R\textsuperscript{2}                & 0.254   & 0.253   & 0.253   & 0.254  & 0.254   \\
\hline

\multicolumn{6}{l}{\textbf{Panel C}. AI-Related Risk}                                 \\
\hline
Dep Var                & \multicolumn{5}{c}{$\textbf{1}\left(\text{\# AI Patent}>0\right)$}                          \\ 
\cline{2-6}
~                      & (1)      & (2)       & (3)      & (4)       & (5)        \\ 
\hline

\textit{ AIRiskSum}    & 4.085***  & ~      & -   & ~   & 1.723***   \\
~             & (5.83) & ~      & ~   & ~   & (2.69)  \\
\textit{ AIRiskAssess}   & ~      & 3.984***  & ~   & -   & 3.405***   \\
~               & ~      & (7.72) & ~   & ~   & (6.96)  \\
\textit{ AIRiskBigram}    & ~      & ~      & -   & -   & -       \\
~               & ~      & ~      & ~   & ~   & ~       \\
Controls        & Yes    & Yes    & ~   & ~   & Yes     \\
Time FE         & Yes    & Yes    & ~   & ~   & Yes     \\
Ind FE         & Yes    & Yes    & ~   & ~   & Yes     \\
N               & 39937  & 39937  & ~   & ~   & 39937   \\
Adjusted R\textsuperscript{2}          & 0.231  & 0.233  & ~   & ~   & 0.233\\
\hline
\end{tabular}
\end{footnotesize}
\end{table}

\clearpage
\newpage

\section*{Table 9. Out-of-Sample Analysis: Economic Outcomes}\phantomsection\label{t9}
\begin{spacing}{0.9}
{\footnotesize\noindent This table repeats the analysis of Tables 7 and 8 with a sample from 2022 to March 2023. We replicate the findings of Table 7 in Panel A and Table 8 in Panel B. We use industry and time fixed effects. All continuous variables are winsorized at 1\% and 99\% levels. Standard errors are clustered at firm-level. ***, **, and * denote significance at 1\%, 5\%, and 10\% levels, respectively.}
\end{spacing}

\begin{table}[hbt!]
\centering
\begin{footnotesize}
\begin{tabular}{p{2.2cm}>{\centering}p{1.9cm}>{\centering}p{1.9cm}>{\centering}p{1.9cm}>{\centering}p{1.9cm}>{\centering}p{1.9cm}>{\centering\arraybackslash}p{1.9cm}} 
\hline
\multicolumn{7}{l}{\textbf{Panel A}. Risk Exposure and Investments}                                 \\ 
\hline
~                       & \multicolumn{6}{c}{Capital Expenditure}                          \\ 
\cline{2-7}
~                       & (1)      & (2)                  & (3)    & (4)                     & (5)    & (6)                   \\ 
\hline

\textit{ PRiskSum }     & -1.838** &           &         &          &        &           \\
\textit{ ~ }            & (-2.50)  &           &         &          &        &           \\
\textit{ PRiskAssess }  &          & -1.533*** &         &          &        &           \\
\textit{ ~ }            &          & (-3.04)   &         &          &        &           \\
\textit{ CRiskSum }     &          &           & -1.304  &          &        &           \\
\textit{ ~ }            &          &           & (-1.16) &          &        &           \\
\textit{ CRiskAssess }  &          &           &         & -1.659** &        &           \\
\textit{ ~ }            &          &           &         & (-2.24)  &        &           \\
\textit{ AIRiskSum }    &          &           &         &          & 2.115* &           \\
\textit{ ~ }            &          &           &         &          & (1.89) &           \\
\textit{ AIRiskAssess } &          &           &         &          &        & 2.604***  \\
\textit{ ~ }            &          &           &         &          &        & (3.25)  \\
Controls                & Yes      & Yes                  & Yes    & Yes                     & Yes    & Yes                   \\
Time FE                 & Yes      & Yes                  & Yes    & Yes                     & Yes    & Yes                   \\
Industry FE             & Yes      & Yes                  & Yes    & Yes                     & Yes    & Yes                   \\

N                       & 9949     & 9949                 & 9949  & 9949                   & 9949  & 9949                 \\
Adjusted R\textsuperscript{2}                  & 0.340    & 0.340                & 0.339  & 0.340                   & 0.340  & 0.340                 \\

\hline
\multicolumn{7}{l}{\textbf{Panel B}. Risk Exposure and Firm Responses}                                 \\ 
\hline
~                       & \multicolumn{2}{c}{$\textbf{1}\left(\text{\$ Lobby Amount}>0\right)$} & \multicolumn{2}{c}{$\textbf{1}\left(\text{\# Green Patent}>0\right)$} & \multicolumn{2}{c}{$\textbf{1}\left(\text{\# AI Patent}>0\right)$}  \\
\cline{2-7}
~                       & (1)      & (2)                  & (3)    & (4)                     & (5)    & (6)                   \\ 
\hline

\textit{ PRiskSum }     &  0.633*& ~                    & ~      & ~                       & ~      & ~                     \\
\textit{ ~ }            &  (1.75) & ~                    & ~      & ~                       & ~      & ~                     \\
\textit{ PRiskAssess }  & ~        & 0.935***            & ~      & ~                       & ~      & ~                     \\
\textit{ ~ }            & ~        & (3.69)             & ~      & ~                       & ~      & ~                     \\
\textit{ CRiskSum }     & ~        & ~                    & 0.457  & ~                       & ~      & ~                     \\
\textit{ ~ }            & ~        & ~                    & (1.31) & ~                       & ~      & ~                     \\
\textit{ CRiskAssess }  & ~        & ~                    & ~      & 0.473**                 & ~      & ~                     \\
\textit{ ~ }            & ~        & ~                    & ~      & (2.49)                  & ~      & ~                     \\
\textit{ AIRiskSum }    & ~        & ~                    & ~      & ~                       & 0.272  & ~                     \\
\textit{ ~ }            & ~        & ~                    & ~      & ~                       & (0.90) & ~                     \\
\textit{ AIRiskAssess } & ~        & ~                    & ~      & ~                       & ~      & 0.547***              \\
\textit{ ~ }            & ~        & ~                    & ~      & ~                       & ~      & (2.89)                \\
Controls                & Yes      & Yes                  & Yes    & Yes                     & Yes    & Yes                   \\
Time FE                 & Yes      & Yes                  & Yes    & Yes                     & Yes    & Yes                   \\
Industry FE             & Yes      & Yes                  & Yes    & Yes                     & Yes    & Yes                   \\

N                       & 11148  & 11148                 & 11148  & 11148                   & 11148  & 11148                 \\
Adjusted R\textsuperscript{2}                  & 0.197    & 0.198                & 0.089  & 0.090                   & 0.067  & 0.068                 \\
\hline
\end{tabular}

\end{footnotesize}
\end{table}

\clearpage
\newpage

\section*{Table 10. Relative Importance of Risk Information}\phantomsection\label{t10}
\begin{spacing}{0.9}
{\footnotesize\noindent This table reports the relative importance of each risk measure over time. We set four-quarter rolling estimation windows. In Panel A, we estimate $Implied\_Volatility_{it}  = \beta RiskAssess_{it} + \gamma \textbf{X}_{it} + \delta_q + \delta_s + \varepsilon_{it}$ separately for $PRiskAssess$, $CRiskAssess$, and $AIRiskAssess$, and report the coefficients and their \textit{t}-values. In Panel B, we estimate $Implied\_Volatility_{it} =  \beta_1 PRiskAssess_{it} + \beta_2 CRiskAssess_{it} + \beta_3 AIRiskAssess_{it}  + \gamma \textbf{X}_{it} + \delta_q + \delta_s + \varepsilon_{it}$ and report $\beta_1$, $\beta_2$, and $\beta_3$, and their corresponding \textit{t}-values. \textit{t}-values are clustered at the firm-level. All continuous variables are winsorized at 1\% and 99\% levels.}
\end{spacing}

\begin{table}[hbt!]
\centering
\begin{footnotesize}
\begin{tabular}{p{2cm}>{\centering}p{2cm}>{\centering}p{1.5cm}>{\centering}p{1.5cm}>{\centering}p{1.5cm}>{\centering}p{1.5cm}>{\centering}p{1.5cm}>{\centering\arraybackslash}p{1.5cm}} 
\hline
\multicolumn{8}{l}{\textbf{Panel A}. Separate Regressions}                                                                                               \\ 
\hline
~      & ~      & \multicolumn{2}{c}{\textit{PRiskAssess ~}} & \multicolumn{2}{c}{\textit{CRiskAssess ~}} & \multicolumn{2}{c}{\textit{AIRiskAssess ~}}  \\ 
\cline{3-8}
~      & ~      & Coeff & t-stat                             & Coeff  & t-stat                            & Coeff  & t-stat                              \\
Start  & End    & (1)   & (2)                                & (3)    & (4)                               & (5)    & (6)                                 \\ 
\hline
2018Q1 & 2018Q4 & 0.374 & 1.72                               & -0.131 & -0.39                             & 0.349  & 0.87                                \\
2018Q2 & 2019Q1 & 0.386 & 1.80                               & 0.180  & 0.52                              & 0.331  & 0.78                                \\
2018Q3 & 2019Q2 & 0.509 & 2.48                               & 0.226  & 0.68                              & -0.002 & -0.00                               \\
2018Q4 & 2019Q3 & 0.438 & 2.05                               & 0.524  & 1.52                              & -0.110 & -0.26                               \\
2019Q1 & 2019Q4 & 0.492 & 2.30                               & 0.519  & 1.50                              & -0.082 & -0.19                               \\
2019Q2 & 2020Q1 & 1.085 & 4.75                               & 2.016  & 5.59                              & -0.596 & -1.26                               \\
2019Q3 & 2020Q2 & 1.360 & 5.52                               & 2.734  & 7.43                              & -1.059 & -2.22                               \\
2019Q4 & 2020Q3 & 1.274 & 5.07                               & 2.627  & 7.30                              & -0.706 & -1.38                               \\
2020Q1 & 2020Q4 & 1.112 & 4.20                               & 2.307  & 6.27                              & -0.625 & -1.24                               \\
2020Q2 & 2021Q1 & 1.020 & 3.97                               & 1.713  & 4.73                              & 0.052  & 0.11                                \\
2020Q3 & 2021Q2 & 0.807 & 3.32                               & 1.042  & 2.86                              & 0.663  & 1.57                                \\
2020Q4 & 2021Q3 & 0.739 & 3.28                               & 0.583  & 1.80                              & 0.431  & 1.08                                \\
2021Q1 & 2021Q4 & 0.698 & 3.10                               & 0.478  & 1.49                              & 0.454  & 1.17                                \\
2021Q2 & 2022Q1 & 0.508 & 2.33                               & 0.501  & 1.65                              & 0.724  & 2.11                                \\
2021Q3 & 2022Q2 & 0.776 & 3.41                               & 1.062  & 3.44                              & 1.106  & 3.08                                \\
2021Q4 & 2022Q3 & 1.148 & 4.61                               & 1.858  & 5.49                              & 1.228  & 3.32                                \\
2022Q1 & 2022Q4 & 1.289 & 4.95                               & 2.108  & 6.02                              & 1.391  & 3.66                                \\
2022Q2 & 2023Q1 & 1.413 & 5.11                               & 2.311  & 5.77                              & 0.996  & 2.76                                \\
\hline
\multicolumn{8}{l}{\textbf{Panel B}. Regressions with Three Risk Measures Together}                                                                      \\ 
\hline
~      & ~      & \multicolumn{2}{c}{\textit{PRiskAssess ~}} & \multicolumn{2}{c}{\textit{CRiskAssess ~}} & \multicolumn{2}{c}{\textit{AIRiskAssess ~}}  \\ 
\cline{3-8}
~      & ~      & Coeff & t-stat                             & Coeff  & t-stat                            & Coeff  & t-stat                              \\
Start  & End    & (1)   & (2)                                & (3)    & (4)                               & (5)    & (6)                                 \\ 
\hline
2018Q1 & 2018Q4 & 0.448 & 2.04                               & -0.414 & -1.24                             & 0.328  & 0.82                                \\
2018Q2 & 2019Q1 & 0.395 & 1.80                               & -0.072 & -0.21                             & 0.290  & 0.68                                \\
2018Q3 & 2019Q2 & 0.525 & 2.49                               & -0.075 & -0.22                             & -0.077 & -0.18                               \\
2018Q4 & 2019Q3 & 0.385 & 1.73                               & 0.292  & 0.82                              & -0.200 & -0.48                               \\
2019Q1 & 2019Q4 & 0.446 & 1.99                               & 0.233  & 0.65                              & -0.200 & -0.47                               \\
2019Q2 & 2020Q1 & 0.640 & 2.63                               & 1.572  & 4.11                              & -0.947 & -2.00                               \\
2019Q3 & 2020Q2 & 0.597 & 2.22                               & 2.331  & 5.78                              & -1.540 & -3.22                               \\
2019Q4 & 2020Q3 & 0.430 & 1.54                               & 2.357  & 5.89                              & -1.192 & -2.31                               \\
2020Q1 & 2020Q4 & 0.343 & 1.18                               & 2.111  & 5.24                              & -1.067 & -2.08                               \\
2020Q2 & 2021Q1 & 0.569 & 2.02                               & 1.350  & 3.40                              & -0.306 & -0.64                               \\
2020Q3 & 2021Q2 & 0.577 & 2.20                               & 0.634  & 1.62                              & 0.410  & 0.96                                \\
2020Q4 & 2021Q3 & 0.686 & 2.77                               & 0.115  & 0.33                              & 0.233  & 0.58                                \\
2021Q1 & 2021Q4 & 0.672 & 2.69                               & 0.022  & 0.06                              & 0.294  & 0.75                                \\
2021Q2 & 2022Q1 & 0.415 & 1.73                               & 0.165  & 0.50                              & 0.607  & 1.74                                \\
2021Q3 & 2022Q2 & 0.500 & 1.99                               & 0.632  & 1.88                              & 0.908  & 2.51                                \\
2021Q4 & 2022Q3 & 0.633 & 2.36                               & 1.327  & 3.73                              & 0.923  & 2.49                                \\
2022Q1 & 2022Q4 & 0.719 & 2.57                               & 1.502  & 4.08                              & 1.047  & 2.76                                \\
2022Q2 & 2023Q1 & 0.864 & 2.94                               & 1.669  & 3.95                              & 0.713  & 2.99                                \\
\hline
\end{tabular}
\end{footnotesize}
\end{table}

\clearpage
\newpage

\section*{Table 11. Asset Pricing on Firm-Level Risk Assessment}\phantomsection\label{t11}
\begin{spacing}{0.9}
{\footnotesize\noindent In Panel A, we present \citet{fama1973risk} regression results and Newey-West \textit{t}-values (with lag=3). We regress monthly stock returns on our firm-level risk exposure measures and the following control variables: stock return for the prior month, stock return for the prior year (skipping the most recent month), log of the market value of equity, log of book-to-market ratio, operating profitability, and investment. We use returns from January 2018 to March 2023. $r_{0,1}$ is the lagged monthly return, $r_{2,12}$ is the lagged yearly return after skipping a month. \textit{Investment} is the assets growth ratio. \textit{Profitability} is (total revenue – cost of goods sold – (sales, administrative expense – R\&D expense)) scaled by total assets. log(\textit{ME}) is the natural logarithm of the market value, and log(\textit{BE/ME}) is the natural logarithm of the book-to-market ratio. We use the natural logarithm of annualized risk assessment measures. Annualized risk assessment is the average value of quarterly risk assessment values. All continuous independent variables are trimmed at 1\% and 99\%. In Panel B, we form portfolios on March 31 of the subsequent year and delete observations with zero annualized risk exposure. We report quintile portfolio alphas using \citet{fama2015five}. We then report monthly high-minus-low alphas and their corresponding t-values. t-values are Newey-West adjusted with a lag 3. *, **, and *** denote statistical significance at 10\%, 5\%, and 1\% levels, respectively.}
\end{spacing}

\begin{table}[hbt!]
\centering
\begin{footnotesize}
\begin{tabular}{p{5cm}>{\centering}p{3.3cm}>{\centering}p{3.3cm}>{\centering\arraybackslash}p{3.3cm}} 
\hline
\multicolumn{4}{l}{\textbf{Panel A}. \citet{fama1973risk} Regressions ~ ~ ~}  \\ 
\hline
Dep Var       & \multicolumn{3}{c}{Return ~ ~}                        \\ 
\cline{2-4}
~             & (1)     & (2)     & (3)                               \\ 
\hline
$r_{0,1}$        & -0.009  & -0.001  & -0.010                            \\
~             & (-0.67) & (-0.67) & (-0.72)                           \\
$r_{2,12}$       & 0.002   & 0.002   & 0.002                             \\
~             & (0.47)  & (0.49)  & (0.33)                            \\
log(\textit{ME})       & -0.000  & -0.000  & -0.000                            \\
~             & (-0.34) & (-0.33) & (-0.32)                           \\
log(\textit{BE}/\textit{ME})    & -0.001  & -0.001  & -0.001                            \\
~             & (-0.54) & (-0.56) & (-0.44)                           \\
\textit{Profitability} & 0.001   & 0.001   & 0.001                             \\
~             & (0.54)  & (0.60)  & (0.62)                            \\
\textit{Investment}    & -0.006  & -0.008  & -0.005                            \\
~             & (-0.65) & (-0.80) & (-0.55)                           \\
\textit{PRiskAssess\textsuperscript{ann}}   & 0.077   & ~       & ~                                 \\
~             & (1.36)  & ~       & ~                                 \\
\textit{CRiskAssess\textsuperscript{ann}}   & ~       & 0.211*  & ~                                 \\
~             & ~       & (1.98)  & ~                                 \\
\textit{AIRiskAssess\textsuperscript{ann}}  & ~       & ~       & 0.317*                            \\
~             & ~       & ~       & (1.72)                            \\
Adjusted R\textsuperscript{2}   & 0.054   & 0.054   & 0.055                             \\
\hline
\multicolumn{4}{l}{\textbf{Panel B}. Single Sorts}  \\ 
\hline
~                & \multicolumn{3}{c}{\citet{fama2015five} Five-Factor Alphas}               \\ 
\cline{2-4}
Sorts On =       & \textit{PRiskAssess\textsuperscript{ann}} & \textit{CRiskAssess\textsuperscript{ann}} & \textit{AIRiskAssess\textsuperscript{ann}}  \\
~                & (1)                  & (2)                  & (3)                    \\ 
\hline
Low              & -0.06                & -0.14                & -0.11                  \\
2                & -0.05                & -0.06                & 0.23                   \\
3                & 0.25                 & 0.08                 & 0.05                   \\
4                & 0.30                 & 0.21                 & 0.16                   \\
High             & 0.38                 & 0.42                 & 0.42                   \\ 
\hline
High – Low       & 0.44                 & 0.56*                & 0.53**                 \\
\textit{t}-value & (1.51)               & (1.90)               & (2.31)                 \\
\hline
\end{tabular}
\end{footnotesize}
\end{table}

\end{document}